\documentclass[prc,
floatfix]{revtex4-2}
\usepackage[utf8]{inputenc}
\usepackage{amsmath}
\usepackage{amsfonts}
\usepackage{amssymb}
\usepackage{graphicx}
\usepackage{fullpage}
\usepackage{mathtools}
\usepackage{bm}
\usepackage{subfig}
\usepackage{xcolor}

\DeclareUnicodeCharacter{2212}{-}

\begin{document}
\title{Uncertainty quantification of transition operators in the empirical shell model}

\author{Jordan M. R. Fox}
\email{jfox@anl.gov}
\affiliation{ Argonne National Laboratory, Lemont, Illinois, USA}
\affiliation{ San Diego State University, San Diego, California, USA}
\author{Calvin W. Johnson}
\email{cjohnson@sdsu.edu}
\affiliation{ San Diego State University, San Diego, California, USA}
\author{Rodrigo Navarro Perez}
\email{rnavarroperez@sdsu.edu}
\affiliation{Mt. San Jacinto College, San Jacinto, California, USA}
\affiliation{ San Diego State University, San Diego, California, USA}

\date{\today}

\begin{abstract}
    While empirical shell model calculations have successfully described low-lying nuclear data for 
decades, only recently has significant effort been made to quantify the uncertainty in such calculations. 
Here we quantify the statistical error in effective parameters for transition operators in 
empirical calculations in the $sd$ ($1s_{1/2}$-$0d_{3/2}$-$0d_{5/2}$) valence space, specifically 
the quenching of Gamow-Teller transitions, effective charges for electric quadrupole ($E2$) transitions, 
and the effective orbital and spin couplings for magnetic dipole ($M1$) transitions and moments.  
We find the quenching factor for Gamow-Teller transitions relative to free-space values is tightly constrained. 
For effective $M1$ couplings, we found isoscalar components more constrained than isovector.  
This detailed quantification of uncertainties, while highly empirical, 
nonetheless is an important step towards interpretation of experiments.
\end{abstract}

\maketitle


\section{Introduction}

Over the last two decades, interest in uncertainty quantification (UQ) has grown rapidly in many sciences, and theoretical nuclear physics is no exception  \cite{dobaczewski2014error,navarro_perez_error_analysis_2015, xu2021bayesian, Lovell_2020, piekarewicz2015information, whitehead2021prediction, drischler2020quantifying, perez2020uncertainty,PhysRevC.96.054316,PhysRevC.98.061301}.
Providing theoretical predictions with error bars that reflect the true limits in our knowledge of a physical system allows for meaningful comparisons between different theoretical models. 
Furthermore, UQ can be considered a driver of the experiment-theory feedback loop. Proper UQ on the theoretical side can be used to identify the experimental measurements that will have the largest impact in reducing such uncertainty. New measurements, with their own error bars, can then be used to improve current theoretical models. This relationship between theory and experiment becomes particularly relevant in the context of upcoming flagship experimental programs like those at the Facility for Rare Isotope Beams (FRIB)~\cite{balantekinNuclearTheoryScience2014}, the Deep Underground Neutrino Experiment (DUNE)~\cite{acciarri2016long}, as well 
as dark matter searches~\cite{undagoitia2015dark} and 
neutrinoless double-beta decay experiments~\cite{engel2017status,dolinski2019neutrinoless}.

The fundamental idea of UQ for theoretical calculations is that no physical theory or model exactly represents nature, and that we can gain insight by studying discrepancy between our simulator and experimental observation. 
We describe an experimental measurement $o$ as 
\begin{equation}
    o(A) = m(A,\theta) + \delta (A,\theta) + \epsilon(A) ,
    \label{eq:uq_model}
\end{equation}
where $m$ is our simulator for the quantity $o$, $A$ denotes attributes of the physical system being modeled, $\theta$ is a set of simulator parameters, $\delta $ is the systematic discrepancy in the simulator, and finally $\epsilon $ is the error in the experiment, including statistical and systematic errors \cite{kennedy2001bayesian,salter_uncertainty_2019}. 
The experimental values $o(A)$ are tabulated with corresponding uncertainties $\epsilon(A)$ being reported by the experimentalist. 
In the present work, $m(A,\theta)$ is the theoretical simulation, the empirical nuclear shell model (including the code implementation, but we ensure code errors are negligible).
The discrepancy $\delta(A,\theta)$ is simply whatever balances out the equation; here, the detailed structure of $\delta (A,\theta)$ is known in principle (e.g. certain \textit{ab initio} models for nuclear structure can explain much discrepancy, e.g.~\cite{PhysRevC.105.034333}), but those calculations can require enormous computational resources. 
Rather than attempting to improve the model by explicitly reducing discrepancy, we want to quantify the uncertainty of model parameters $\theta$. Because of this, we consider the total error $\delta (A,\theta) + \epsilon(A) = o(A) - m(A,\theta)$, and uncertainty on $\theta$ can be determined by studying this relationship for many different cases or systems $A$.

The  UQ endeavor includes not only computing error bars for theoretical predictions, but also studying correlations between our variables, whether those be solutions (wavefunctions), coupling constants, observables, etc.
Bayesian statistical analyses \cite{sivia_bayes}, which are a natural fit for probabilistic modeling of theory, have become popular in the nuclear science community. 
This formulation allows us to go beyond the standard least-squares fitting procedure and to describe non-Gaussian distributions of the variables in question, which is important for the calculations described in Section \ref{sec:parameter_estimation}.
Significant advances in computational techniques and tools like Markov Chain Monte Carlo (MCMC) \cite{mcmc_handbook}, Hamiltonian MC \cite{hmc}, NUTS \cite{nuts}, and the \texttt{emcee} library \cite{emcee_paper} can help achieve efficient and accurate evaluation of probability distributions. Furthermore, the development of emulator models (eigenvector continuation \cite{frame_ec,konig_ec}, Gaussian processes \cite{rasmussen_gp}, and even neural networks \cite{goodfellow_deeplearning, neufcourt_bayesian_extrapolation}) allows researchers to study perturbations to complicated models, trading some accuracy for far less computation.  

In this work we apply UQ techniques to nuclear shell model calculations to analyze variability of coupling constants with respect to a particular nuclear force model, which in turn requires a statistical description of individual transition matrix elements.   
In particular we use MCMC and the implementation of the \texttt{emcee} library for sampling probability distributions. Future research on shell-model UQ should also investigate other methods, especially emulators for calculations which require significant computational resources. 
%
%
%
After a brief description of the empirical nuclear shell model in Section \ref{sec:shell_model}, we present the statistical treatment of model parameters in Section \ref{sec:sensitivity_analysis}, and how to construct the distributions of observables and operator parameters in Section \ref{sec:parameter_estimation}. In Section \ref{sec:results} we show results of parameter estimation for transition operator parameters. Appendix \ref{appendix:hessian_comparisons} has comparisons of parameter sensitivity for all observables in question; 
this specifically allows us to identify key matrix elements which could be fitted simultaneously to 
both energy levels and transition strengths.
We include supplemental materials with data sets and additional plots.

\section{The empirical configuration-interaction shell model} \label{sec:shell_model}

A long-standing approach to low-energy nuclear structure is 
the configuration-interaction shell model~\cite{BG77,br88,ca05,suhonen2007nucleons}, whereby one solves 
the time-independent Schr\"odinger equation, $\hat{H}|\psi \rangle = E|\psi \rangle$, by expanding in a basis and converting to a 
matrix eigenvalue problem.  Here we use an orthonormal basis,
 antisymmetrized products of single-particle states, that is,
the occupation representation of Slater determinants. 
Such a framework is relatively transparent and in principle 
exact; in addition, relevant to this work, generating many low-lying excited states is not much more difficult than generating the ground state. The main drawback is that one needs a large number of basis 
states, a demand which grows exponentially with the single-particle space 
and the number of particles.

Nuclear configuration-interaction shell model calculations typically fall into two categories: empirical (a.k.a. phenomenological or effective) and \textit{ab initio}. The latter, such as the no-core shell model \cite{navratil2000large,barrett2013ab},   uses Hamiltonians derived 
primarily from two- and few-nucleon data, often in the framework of
chiral effective field theory~\cite{PhysRevC.49.2932}; because they
work in large single-particle spaces and converge 
as the model space increases, such calculations 
are considered more robust than empirical calculations, but are limited to $p$-shell nuclides and  the lightest $sd$-shell nuclides.  Part of the motivation for such 
calculations is that one can identify the errors in the theory 
 from leaving out higher order 
 terms~ \cite{wendt2014uncertainty,furnstahl2015recipe,carlsson2016uncertainty,perez2016uncertainty,PhysRevC.100.044001,wesolowski2019exploring}, but in practice  
 UQ in \textit{ab initio} many-body calculations is far from trivial. 
 
 Empirical calculations \cite{BG77,br88,ca05,suhonen2007nucleons} on the other hand have a long, rich history of applications to heavier nuclei; the trade-off is that their  interaction matrix elements are restricted to a fixed valence space
 and are fit to reproduce a set of experimental observations, usually 
 binding and excitation energies, of nuclides within the valence space. Hence if one expands the 
 model space, the matrix elements of necessity must be refit. 
 A promising alternative is so-called microscopic effective interactions ~\cite{stroberg2019nonempirical} wherein \textit{ab initio} methods are used in a fixed model space, but these are still very much under development. 
 
A nuclear Hamiltonian with one- and two-body components can be expressed in the formalism of second-quantization:
\begin{equation}
    \label{ham}
    \hat{{H}} = \sum_{rs} T_{rs} \hat{a}^\dagger_r \hat{a}_s + \frac{1}{4}
    \sum_{rstu} V_{rstu}\hat{a}^\dagger_r \hat{a}^\dagger_s \hat{a}_u \hat{a}_t 
\end{equation}
Here,  $\hat{a}^\dagger_r, \hat{a}_s$ are particle creation/annihilation operators,  while the indices $r,s,$ etc.,  label single-particle states 
with good angular momentum quantum numbers such as orbital angular 
momentum $l$, total angular momentum $j$, and $z$-component $m$. The
$T_{rs}$ are one-body matrix elements, including kinetic energy and 
interaction with a frozen core and typically diagonal, $r=s$, in which case they are called single-particle energies; and  the $V_{rs,tu}$ are two-body matrix elements. The matrix 
elements themselves conserve angular momentum and, in some cases, isospin.  Thus the 
$V_{rs,tu}$ are not all independent. Instead, the independent matrix elements are written 
 $V_{JT}(ab,cd) = \langle ab; JT| \hat{V} | cd; JT\rangle$, where $\hat{V}$ is the nuclear two-body interaction and $| ab; JT \rangle$ 
is a normalized two-body state with nucleons in single-particle orbits (that is, with 
quantum numbers $l$ and $j$ but not $m$) labeled by $a,b$ coupled
up to total angular momentum $J$ and total isospin $T$; the $V_{rs,tu}$ in Eq.~(\ref{ham}) 
are the uncoupled matrix elements found by expanding with Clebsch-Gordan coefficients. 
(Note that even here, there are relations among these matrix elements, for example 
$V_{JT}(ab,cd)$ is related to $V_{JT}(ba,cd)$. While the details are important for calculations
and well-known to experts~\cite{br88,ca05}, they are not deeply relevant to our arguments here.) 
Similarly we have single-particle energies $\epsilon_a$ for each orbital.

The independent coefficients $\{ \epsilon_a, V_{JT}(ab,cd) \} = \{ \lambda_i \} =  \bm{\lambda}$ are a set of real numbers which parameterize the interaction, that is, we write
\begin{equation}
    \label{ham2}
    \hat{H} = \sum_i \lambda_i \hat{o}_i
\end{equation}
where $\hat{o}_i$ is the appropriate operator. For empirical calculations the values of the $\{ \lambda_i \}$ 
are fitted to reproduce some observables, almost always energies. 

Here we note an important source of 
model error.  Although the interaction parameters often originally arise from some known translationally-invariant 
(i.e., relative coordinates) representation, and the matrix elements calculated as integrals 
using some choice of single-particle wave functions, 
the matrix elements are directly adjusted in the laboratory frame 
to fit data~\cite{BG77,br88}, thus severing any 
explicit connection to an underlying potential or Lagrangian. (This differs from \textit{ab 
initio} calculations, where one may carry out specific transformations of the 
interaction~\cite{PhysRevC.75.061001,hergert2016medium}
but never loses the thread back to the ``original'' interaction.) Thus, arguably, one can no 
longer even fix with certainty the single-particle basis, specifically the radial part of 
the wave functions, from which the many-body states are built.  This has consequences 
when computing observables, such as radii or electromagnetic transitions, for which 
the single-particle wave functions are a key input. 
A frequent choice in the literature is to assume harmonic oscillator basis states, but this is 
mostly out of simplicity and convenience: there is a single parameter to choose. 

To solve the many-body problem, we use a  configuration-interaction code, {\sc  Bigstick } \cite{BIGSTICK,johnson2018bigstick}, to generate from the single-particle energies 
and two-body matrix elements the many-body Hamiltonian 
and solve for extremal eigenvectors and eigenvalues using the Lanczos algorithm.
{\sc Bigstick} uses an $M$-scheme basis, which means all basis vectors (and thus, all solutions) share the same value for the $z$-component of total angular momentum, $M$ or $J_z$. Because the matrix elements are read in as an external file, {\sc Bigstick} 
does not depend upon a particular choice of single-particle basis.
Any other nuclear configuration-interaction shell model code 
will yield the same results from the same input files.

Our interaction of interest is a widely-used, `gold standard' empirical shell-model interaction, Brown and Richter's universal $sd$-shell interaction, version B, or USDB~\cite{brown_usd_hams}, a set of 66 parameters fit to energy data for a number of nuclei between $^{16}\text{O}$ and $^{40}\text{Ca}$. In previous work \cite{fox_usdb} we performed a sensitivity analysis of USDB, which gives a probability distribution $P(\bm{\lambda})$ from which the interaction is sampled. An important nuance in using the USDB parameters is that while the single-particle energies are fixed, the two-body matrix elements are scaled by $(A_0/A)^{0.3}$, where $A$ is the mass number of the nucleus, and $A_0$ is a reference value, here $= A_\text{core} + 2 = 18$. We account for this by 
modifying the Hamiltonian expression $\hat{H} = \sum_{i} \lambda_i (A_0/A)^{0.3}\hat{\cal O}_i$ (but only for the two-body matrix elements), so that we implicitly varied the parameters fixed at $A=18$.

Eigenstates of $\hat{H}$ give us many-body wavefunctions: $|\psi\rangle$; from these we measure transition matrix elements which depend on different initial and final states. The result is a number related to the probability of state $|\psi_i\rangle$ being transformed to state $|\psi_f\rangle$ by an operator $\hat{O}$ representing 
coupling to an external field:
\begin{equation}
    \label{eq:transition_M}
    M_{if}(\hat{O}) = \langle \psi_f | \hat{O} | \psi_i \rangle 
\end{equation}
The wavefunctions, being solutions to the eigenvalue problem, depend on $\bm{\lambda}$, and we assign the label $\theta$ to any parameters in the operator.
\begin{equation}
    M_{if}(\hat{O};\bm{\lambda},\theta) = \langle \psi_f(\bm{\lambda}) | \hat{O}(\theta) | \psi_i(\bm{\lambda}) \rangle
    \label{eq:tme}
\end{equation}
The experimental transition rate is  proportional to $|M_{if}|^2.$ 
We leave out the technical elements of computing  transition matrix elements,  detailed in the literature~\cite{BG77,br88,ca05,suhonen2007nucleons}.  The most important point is that we consider standard 
one-body transition operators easily represented through second quantization:
\begin{equation}
    \hat{O}(\theta) = \sum_{rs} \omega_{rs}(\theta) \hat{a}_r^\dagger 
    \hat{a}_s,
\end{equation}
where, again, the dependence of the coefficients $\omega_{rs}$ upon
the effective parameters $\theta$ is well-known.

Hence, we have the computational pipeline of nuclear observables: $\bm{\lambda}$ determines the Hamiltonian $\rightarrow$ eigenstates of the Hamiltonian determine wavefunctions $\rightarrow$ matrix 
elements of transition operators between those eigenstates determine transition probabilities. Thus, we pursue two questions. First, given $P(\bm{\lambda})$, how does $M_{if}(\hat{O};\bm{\lambda},\theta)$ behave? And second, given $P(\bm{\lambda})$ and some experimental observations $O$, what is our posterior distribution $P(\theta | O)$?  In many cases, it is useful to produce a covariance matrix for $\theta$, which describes sensitivity of $\theta$ with respect to $O(\bm{\lambda},\theta)$ about an optimal value.

\section{Sensitivity analysis of the interaction}\label{sec:sensitivity_analysis}

Our prior sensitivity analysis in \cite{fox_usdb} assumed that 
the Hamiltonian parameters (matrix elements) 
$\bm{\lambda}$ follow a normal distribution; thus their variability is fully described by a covariance matrix $C_\lambda$. The covariance is the inverse of the Hessian $H_\lambda$ (not to be confused with the Hamiltonian $\hat{H}$), that is, the second derivatives of energy errors with respect to parameters $\bm{\lambda}$ evaluated with the parameters that minimize the error, $\bm{\lambda}^*$. (The notation $\bm{\lambda}^*$ for parameters at the minimum, while not common in nuclear physics, is used in uncertainty quantification \cite{ghanem2017handbook}.) We arrived at two important conclusions: (1) the Hessian matrix can be well approximated by a simpler calculation depending on first derivatives alone, and (2) we confirmed previous findings (e.g., \cite{brown_usd_hams}) that the parameter space is dominated by a small number of principal components, specifically that only 10 of the total 66 linear combinations of parameters account for 99.9\% of the Hamiltonian. (The first most important linear combination carries 95\%.) 

The Hessian matrix is the second derivative of the $\chi^2$ function about the optimal model parameters $\bm {\lambda}^*$:
\begin{equation}
    \left[ H_\lambda \right]_{ij} = \left. \frac{1}{2} \frac{\partial^2}{ \partial \lambda_i \partial \lambda_j}    \chi^2(\bm{\lambda}) \right \rvert_{\bm{\lambda} = \bm{\lambda}^*}
    \label{hessian}
\end{equation}
where 
\begin{equation} \label{eq:chisq_energy}
    \chi^2(\bm{\lambda}) = \sum_{\alpha}^{N_d} \frac{ ( E^{\text{exp}}_\alpha - E^{\text{th}}_\alpha(\bm{\lambda}) )^2 }{ \sigma(E^{\text{exp}}_\alpha)^2 + \sigma(E^{\text{th}})^2 } = \bm{e}^T_E(\bm{\lambda}) C_E^{-1} \bm{e}_E(\bm{\lambda}).
\end{equation}
Here $\bm{e}_E(\bm{\lambda})$ is the energy error vector, $C_E$ is a matrix with energy variances along the diagonal, and $N_d$ is the number of data points. The two contributions to variances are the experimental variance $\sigma(E_\alpha^\mathrm{exp})^2$ and the \textit{a priori} theoretical variance $\sigma(E^\mathrm{th})^2$ which is the same for all observations (hence no $\alpha$ subscript). The latter is tuned such that the $\chi^2$ per degree of freedom,
\begin{equation}\label{eq:reduced_chisq_energy}
    \chi^2_\nu(\bm{\lambda}) = \frac{\chi^2(\bm{\lambda})}{\nu} = \frac{\chi^2(\bm{\lambda})}{N_d - N_p} 
\end{equation}
(a.k.a. the reduced $\chi^2$) is 1. In doing so, we ensure the Hessian matrix be scaled appropriately, so the inverse may be interpreted as covariance. This procedure for determining the ``static'' theoretical uncertainty estimate is the same for the later analyses of transition strengths. 

We found the Hessian for interaction parameters is well approximated by $H_\lambda \approx A_\lambda \equiv J_{E(\lambda)}^T C_E^{-1} J_{E(\lambda)} $ where $\left[ J_{E(\lambda)} \right]_{i \alpha } = \partial E^\text{th}_\alpha (\bm{\lambda}) / \partial \lambda_i$ is the Jacobian matrix, which is computed easily using the Feynman-Hellman theorem~\cite{hellman1937einfuhrung,PhysRev.56.340}:   $\left[ J_{E(\lambda)} \right]_{i \alpha } = \langle \alpha | \hat{\cal O}_i | \alpha \rangle $, an expectation value of eigenstates $\hat{H}(\bm{\lambda}) |\alpha \rangle = E^\text{th}_\alpha (\bm{\lambda}) | \alpha \rangle$. 
This approximation replaces an $O(N_p^2)$ calculation with an $O(N_p)$ calculation while only introducing small errors.  

We thus describe the distribution for interaction parameters $\bm{\lambda}$ as normal with mean at the fitted USDB values and with the covariance given by our approximation:

\begin{equation}
    \bm{\lambda} \sim \mathcal{N} \left( \text{mean} = \bm{\lambda}_\text{USDB} , \text{cov} = [J^T C_E^{-1} J]^{-1} \right)  \equiv P(\bm{\lambda})
    \label{usdb_dist}
\end{equation}

We use this description to propagate uncertainty of $\bm{\lambda}$ to calculations of observables. Implicit in this is the assumption that the covariance of parameters with respect to energies is the same as their covariance with respect to the other observables in question. Our investigation shows that, while the overall scale of covariances may differ, the general structure of the covariance matrices tends to be preserved. We make this approximation in this work for convenience, since computing covariance matrices for other observables would require finite difference estimation. For more information on covariance matrix results see Appendix~\ref{appendix:hessian_comparisons} and the supplemental materials.

\section{Bayesian parameter estimation} \label{sec:parameter_estimation}

Bayesian data analysis is so named because it makes use of \textit{Bayes' rule} to create statistical models:
\begin{equation}\label{eq:bayes}
    P(\theta | y) = \frac{P(y | \theta) P(\theta)}{P(y)}
\end{equation}
The \textit{posterior} $P(\theta | y)$ is the ultimate goal of our analysis: a distribution describing the quantity of interest $\theta$ given observations $y$. The \textit{likelihood} $P(y | \theta)$ describes the converse of the posterior: how observations $y$ behave given $\theta$. The \textit{prior} $P(\theta)$ is the probability distribution of $\theta$ according to our prior knowledge. Finally, the \textit{evidence} $P(y)$ describes any bias to observations $y$, but will only contribute an overall normalization to the posterior.

Following typical Bayesian data analysis, $\theta$ stands for any parameter or vector of parameters we are modeling;  here those parameters are the coupling constants (and, for electric quadrupole transitions, the basis harmonic oscillator length parameter) appearing in transition operators. This should not be confused with the vector of Hamiltonian parameters $\bm{\lambda}$, which is well described by a Gaussian approximation, while $\theta$ will be evaluated by MCMC. To describe $\theta$, we ultimately must compute the posterior distribution with respect to experimental observations $O$, $P(\theta|O)$.
We take several mathematical steps to put the posterior $P(\theta|O)$ into a form we can compute. First, we introduce the interaction parameters as a marginal variable (that is, a new variable being integrated over) in $\mathbb{R}^{N_p}$, where $N_p$ is the number of parameters (here, 66):
\begin{equation}
    P(\theta|O) = \int_{\mathbb{R}^{N_p}}   P(\theta, \bm{\lambda} | O)  \:d\bm{\lambda}
\end{equation}
By the chain rule of probabilities, we can reinterpret $\bm{\lambda}$ as a conditional variable if we also insert its prior in the integrand.
\begin{equation}
    \int_{\mathbb{R}^{N_p}} P(\theta, \bm{\lambda} | O) \: d\bm{\lambda}  = \int_{\mathbb{R}^{N_p}} P(\theta| O,\bm{\lambda}) P(\bm{\lambda}) \:d\bm{\lambda}
\end{equation}
Since we can easily sample the distribution $P(\bm{\lambda})$ via Eq.~(\ref{usdb_dist}), we approximate the integral over  $ P(\theta| O,\bm{\lambda}) P(\bm{\lambda}) $ as an average of $P(\theta| O,\bm{\lambda}_k)$ for $N_s$ interaction samples $\bm{\lambda}_k$.
\begin{equation}
     \int_{\mathbb{R}^{N_p}} P(\theta| O,\bm{\lambda}) P(\bm{\lambda}) \: d\bm{\lambda} \approx \frac{1}{N_s} \sum_{\bm{\lambda}_k \sim P(\bm{\lambda})} P(\theta|  O, \bm{\lambda}_k)  
\end{equation}
%
The integral is better approximated as $N_s \rightarrow \infty$. By the central limit theorem, the sum converges to the integral with errors on the order of $1/\sqrt{N_s}$; to keep this error less than 2\%, we use at least 2,500 samples for each calculation. 

Next, we apply Bayes' rule to the summand. 
As mentioned above, the probability distribution of experimental observations, $P(O)$, often referred to as the ``evidence'', only contributes to overall normalization of the posterior.
We ignore it and the factor of $1/N_s$ since the posterior need not be normalized in order to model $\theta$.
\begin{equation} \label{eq:sum_of_likelihoods}
     \frac{1}{N_s} \sum_{\bm{\lambda}_k \sim P(\bm{\lambda})} P(\theta|  O, \bm{\lambda}_k) 
     = \frac{1}{N_s} \sum_{\bm{\lambda}_k \sim P(\bm{\lambda})} \frac{P(O| \theta , \bm{\lambda}_k) P(\theta)}{P(O)} 
     \propto P(\theta) 
    \sum_{\bm{\lambda}_k \sim P(\bm{\lambda})} P(O| \theta , \bm{\lambda}_k )
\end{equation}
 The right-hand side of this equation we can compute, and the likelihood is:
\begin{equation}\label{eq:one_likelihood}
    P(O| \theta , \bm{\lambda}_k ) = \exp \left[ - \frac{1}{2}\chi_O^2(\theta,\bm{\lambda}_k) \right]
\end{equation}
 The function $\chi_O^2(\theta,\bm{\lambda}_k)$ is similar to Eq.~(\ref{eq:chisq_energy}), but is defined for the observables $O$ instead of energies. 
\begin{equation}\label{eq:chi_squared}
    \chi^2(\theta,\bm{\lambda}_k) = \bm{e}^T_O(\theta,\bm{\lambda}_k) C_O^{-1} \bm{e}_O(\theta,\bm{\lambda}_k),
\end{equation}    
where $[\bm{e}_O(\theta,\bm{\lambda}_k)]_\alpha = (  O_\alpha^\mathrm{exp} - O_\alpha^\mathrm{th}(\theta,\bm{\lambda}_k) )$ is the error vector and $ [C_O]_{\alpha\alpha} = \sigma(O_\alpha^\mathrm{exp})^2 + \sigma(O^\mathrm{th})^2$ is the matrix with total variances along the diagonal. Here, the \textit{a priori} theoretical uncertainty $\sigma(O^\mathrm{th})$ is determined in the same way as for Eq.~(\ref{eq:reduced_chisq_energy}), by setting the reduced $\chi^2$ involving $\theta$,
\begin{equation}\label{eq:reduced_chisq_trans}
    \chi^2_{\nu,O}(\theta, \bm{\lambda}_k) = \frac{\chi_O^2(\theta, \bm{\lambda}_k)}{\nu} = \frac{\chi_O^2(\theta,\bm{\lambda}_k)}{N_d - N_p},
\end{equation}
to unity and evaluating at $\bm{\lambda}_\text{USDB}$, which also ensures the Hessian with respect to $O$ has proper scaling, $H_O = C_O^{-1}$. Here, $N_d$ is the number of observations $O$ and $N_p$ is the number of total parameters in $\theta$ and $\bm{\lambda}$.

For simplicity of the calculation, $\sigma(O^\mathrm{th})$ is computed once using a reasonable \textit{a priori} choice of parameters, and does not change with samples of $\theta$ and $\bm{\lambda}$.
Finally, we have an expression for the posterior distribution $\theta$ in terms of things we can compute:
\begin{equation}\label{eq:posterior}
    P(\theta|O) \propto P(\theta) \sum_{\bm{\lambda}_k \sim P(\bm{\lambda})} \exp \left[ - \frac{1}{2}\bm{e}^T_O(\theta,\bm{\lambda}_k) C_O^{-1} \bm{e}_O(\theta,\bm{\lambda}_k) \right]
\end{equation}

Each transition operator has unique parameters $\theta$ and observations $O$, but the general process of describing $P(\theta|O)$ is the same for any observable. We decide on the prior $P(\theta)$, construct the expression in Eq.~(\ref{eq:posterior}), and measure it. Since contributions to the likelihood in Eq.~(\ref{eq:one_likelihood}) are highly non-Gaussian, we evaluate using MCMC, the affine-invariant ensemble sampler from  \texttt{emcee}~\cite{emcee_paper}. However, by the central limit theorem the sum over $P(O| \theta , \bm{\lambda}_k )$ approaches a Gaussian as the number of observations becomes large, meaning $\theta$ can ultimately be well described by mean and covariance alone.
 
Due to our frequentist approximation of the total likelihood function, the above procedure might be called ``pseudo-Bayesian''. A fully Bayesian analysis, rather than summing over many likelihoods, would probably define  likelihood as a function of  $(\theta,\bm{\lambda})$ together and construct the posterior accordingly: $P(\theta)P(\bm{\lambda})P(O|\theta,\bm{\lambda})$. Indeed, doing so would elucidate correlations between the Hamiltonian matrix elements and operator parameters. We choose to construct the approximate likelihood from many samples of $\bm{\lambda}$ mainly due to practical challenges. In order to evaluate the likelihood function, which contains a sum over observables in many nuclides, perturbations to the Hamiltonian must not result in the relevant eigenstates vanishing or otherwise being too difficult to track. A single evaluation of the likelihood requires at least dozens of individual shell model calculations, perhaps even hundreds in extreme cases. Sampling according to our prior PCA makes it more likely that these calculations produce a sensible answer for every observable. Our codes compute wavefunction overlaps upon each iteration to help track eigenstates; this grants some robustness but it is not perfect. A potential solution to this problem is using eigenvector continuation, which recently has been applied to shell model calculations \cite{yoshida_eigenvector_continuation}, but that is beyond the scope of this paper.
 
\section{Results}\label{sec:results}

Here we present the results of uncertainty quantification for Gamow-Teller (GT) transitions, electric quadrupole ($E2$) transitions, and magnetic dipole ($M1$) transitions and moments. 
The analyses are presented in order of increasing difficulty:  GT has only one parameter and thus is the simplest;  $E2$ has three but only two are independent; $M1$ has four parameters. 
Most of our experimental observations $O$ are reduced transition strengths, $B(\hat{O}) = |M(\hat{O})|^2/(2 J_i +1)$, where $M(\hat{O})$ is the  reduced~\cite{edmonds1996angular} matrix element for transition operator $\hat{O}$, and $J_i$ is the initial angular momentum. 
The $M1$ observations include dipole moments as well.
Observations outside the range of model validity were excluded, including isotopes which have no protons or neutrons (or proton/neutron holes) in the valence space.
For electric quadrupole ($E2$) and magnetic dipole ($M1$) transitions, we converted to Weisskopf units to get a sense of their strength with respect to single particle estimates, then truncated datasets to exclude extreme cases. $E2$ transitions with strength $<0.1\times$ and $>150\times$ the Weisskopf single particle estimate were excluded. This shrunk the total $E2$ data set from 544 to 483 transitions. 
For $M1$, which tend to be smaller on average, we left out those with $<0.01\times$ the Weisskopf estimate. We also dropped some $M1$ transitions involving very excited states, in particular if either state is above 6 excitations for a particular total angular momentum $J$. This altogether shrunk our $M1$ data set from 575 to 281 transitions and from 73 to 72 moments.  

For each transition operator, we present an \textit{a priori} theoretical uncertainty: it is a fixed estimate of theoretical error based on setting the reduced $\chi^2$ to one using a \textit{prior} parameter estimate. These provide a useful prediction of theory error, but they can be sensitive to choices of the UQ analysis, prior parameterizations, data sets, etc. 
The study of observables using USD Hamiltonians by Richter, Mkhize, and Brown (2008) \cite{richter_sd_obs} serves as our primary source for prior information on parameter uncertainties. 

Exact probability distributions mentioned are either uniform on a closed interval $[a,b]$, written $U_{[a,b]}$, or normal with mean $\mu$ and standard deviation $\sigma$, written $\mathcal{N}(\mu,\sigma)$.

\subsection{Gamow-Teller transitions} 


Both the vector and axial-vector weak couplings, $g_V, g_A$ respectively, have been measured from the $\beta$-decay of a 
free neutron, with $|g_A/g_V| \approx 1.28$ \cite{richter_sd_obs}. 
Empirical shell-model calculations of allowed transitions, specifically Gamow-Teller reduced transition matrix elements
\begin{equation}
M_{if}(\text{GT})= \left \langle J_f \left | \left | 
g^\mathrm{eff}_A \vec{\sigma} \tau_\pm \right | \right | J_i  \right \rangle,
\end{equation}
when compared to experiment, consistently lead to a quenched coupling, $g_A^\text{eff} = Qg_A$~\cite{br88,ca05};  $Q$ is called the \textit{quenching factor}. Recent work in \textit{ab initio} calculations have shown quenching can largely be accounted for by including physics beyond capabilities of the effective shell model, including two-body currents, sometimes interpreted 
as meson exchange, and long-range energy correlations~\cite{gysbers2019discrepancy}.

\begin{figure}
    \includegraphics[scale=0.5,clip]{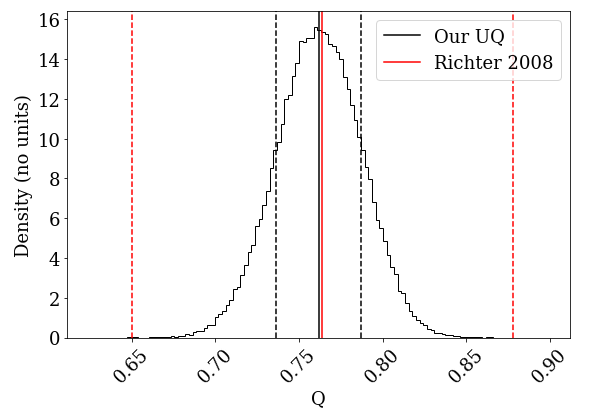} 
    \caption{Histogram of the Gamow-Teller quenching factor $Q$ via Monte Carlo; the posterior is Gaussian, $Q = 0.77 \pm 0.013$. This is more tightly constrained than the estimate from \cite{richter_sd_obs}, $Q = 0.764 \pm 0.114$.}
    \label{fig:Q_hist}
\end{figure}

Using 185 low-lying $\beta^+,\beta^-,$ and electron-capture transitions, we assign an \textit{a priori} theoretical uncertainty in the (dimensionless) $B(\text{GT})$ to be 0.30, based on setting the reduced $\chi^2$ to 1. 
The only \textit{a priori} assumption in our UQ is that $0.5< g_A^\text{eff}/g_A < 1.0$ so our prior is set to a uniform distribution within those bounds: $P(Q) = U_{[0.5,1.0]}$.   Fig.~\ref{fig:Q_hist} show our derived posterior, which is Gaussian with 
 $Q = 0.762 \pm 0.025$.  
This is more tightly constrained than the estimate from Richter et al. 2008, $Q = 0.764 \pm 0.114$ \cite{richter_sd_obs}, but our result is consistent with their conclusion that quenching of $g_A$ in empirical calculations is robust and independent of the Hamiltonian.

\begin{figure}
    \includegraphics[scale=0.4]{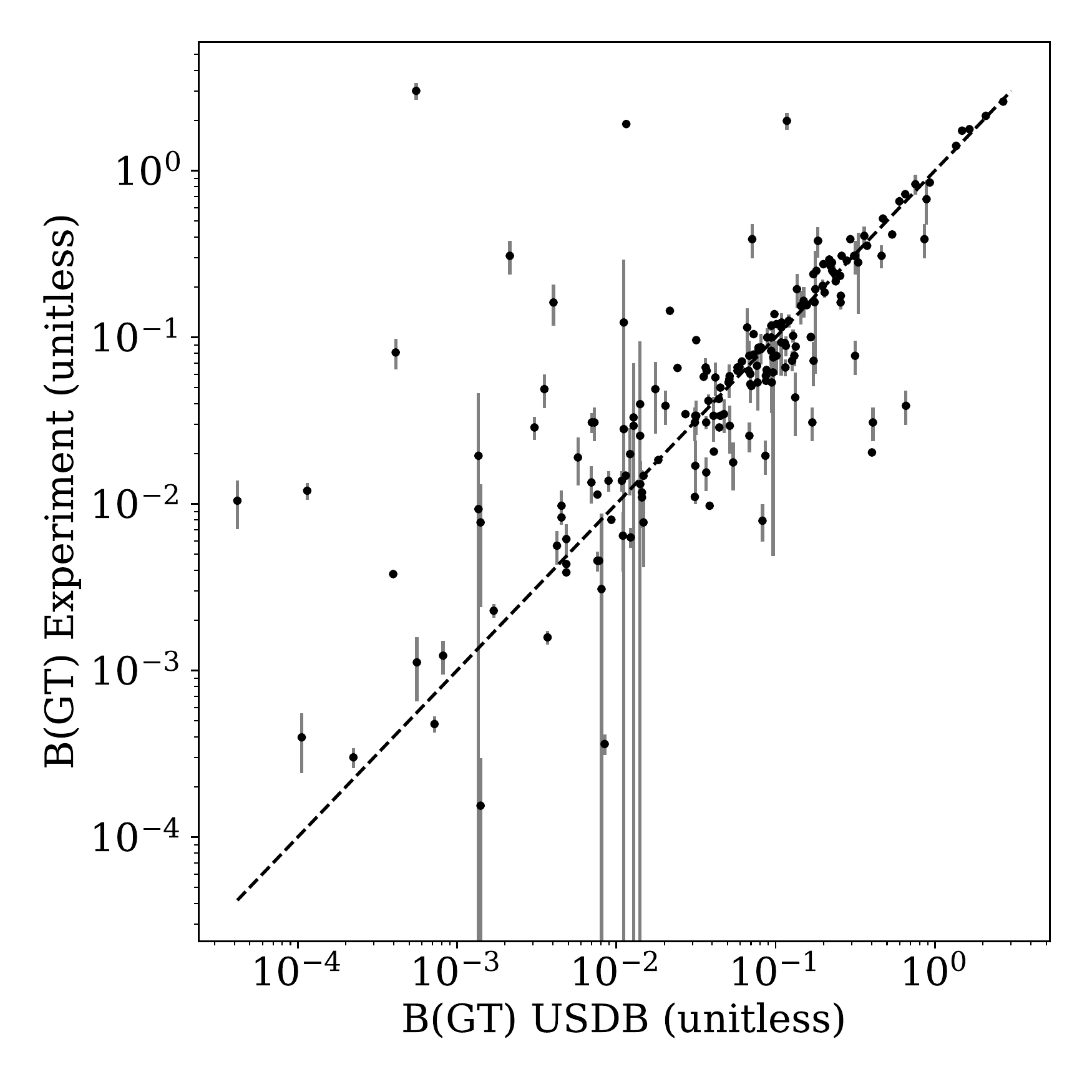} 
    \caption{Gamow-Teller transition strengths computed with the USDB interaction compared with experimental data with experimental error bars and using a quenching factor of $Q=0.77$.}
    \label{fig:GT_scatter}
\end{figure}

Figure ~\ref{fig:GT_scatter} shows a comparison of experimental Gamow-Teller transitions to the USDB calculations using the maximum-likelihood estimate of $Q$. 
The three points at the top have particularly large errors, with experiment reporting larger strengths by several orders of magnitude. These are 
\begin{equation*}
\begin{split}
    ^{31}\text{Cl} \left( 3/2^+_1 \right) & \rightarrow \; ^{31}\text{S} \left( 3/2^+_4 \right)\\
    ^{35}\text{K} \left( 3/2^+_1 \right) & \rightarrow \;^{35}\text{Ar} \left( 3/2^+_3 \right)\\
    ^{36}\text{K} \left( 2^+_1 \right) & \rightarrow \;^{36}\text{Ar} \left( 2^+_4 \right),
\end{split}
\end{equation*}
all of which have a final state with no change in total angular momentum and excitation energies over 6 MeV. For Potassium isotopes, errors could be attributed to truncations of the model space. One can also speculate that, due to the excited final states, the final wavefunctions may be missing structure that is important for the Gamow-Teller. Such speculation may be resolved by performing the same calculations in larger model spaces, but that is beyond our present scope.  

\subsection{Electric quadrupole transitions}


The electric quadrupole ($E2$) reduced transition matrix element is 
\begin{equation}
    M_{if}(E2) = \langle J_f || \left[ e_p E2_p + e_n E2_n \right] || J_i \rangle
\end{equation}
where $E2 := r^2 Y_2^m$ and $e_p, e_n$ are the effective charges for protons and neutrons, respectively. 
The parameters of interest are the effective proton and neutron charges $\theta = (e_p , e_n )$.
Because we assume harmonic oscillator single-particle wave functions, the matrix elements of $r^2$ are proportional to $b^2$, where $b$ is the oscillator length.
Oscillator length is strongly correlated with radii; we use a global fit $b^2 = 0.9 A^{1/3} + 0.7$, known as  Blomqvist-Molinari formula.
It should be noted that the Blomqvist-Molinari formula is fitted to charge radii far beyond just $sd$-shell nuclides and does not include an error estimate. One could instead use either experimental radii or radii 
from, for example, density-functional theory, in order to fix the oscillator parameter, but both of these are beyond the scope of this work.

First, we assign prior distributions for the parameters.
Effective charges for $E2$ calculations take values of $e_p \in (1,2)$ and $e_n \in (0,1)$. Previous work by Richter \textit{et al.} using the USDB interaction found optimal values $e_p=1.36(5)$ and $e_n=0.45(5)$~\cite{richter_sd_obs}.  
These estimates are used to compute our \textit{a priori} theoretical uncertainty estimate of 4.4 Weisskopf units,  based on setting the reduced $\chi^2$ to unity. 

Whether we use a uniform prior, $P(e_p) = U_{[1,2]}$ and $P(e_n) = U_{[0,1]}$, or a Gaussian prior centered on previous optimal results, $P(e_p) = \mathcal{N}(\mu=1.36,\sigma=1)$ and $P(e_p) = \mathcal{N}(\mu=0.45,\sigma=1)$, the resulting posteriors for effective charges are almost identical (including reducing the prior standard deviations down to $0.5$). The assigned uncertainties in ~\cite{richter_sd_obs} are $\sigma=0.05$ for both proton and neutron, too small to construct a sensible prior with, and thus the resulting posterior is constrained mainly by the likelihood.
 
\begin{figure}
\includegraphics[scale=0.5]{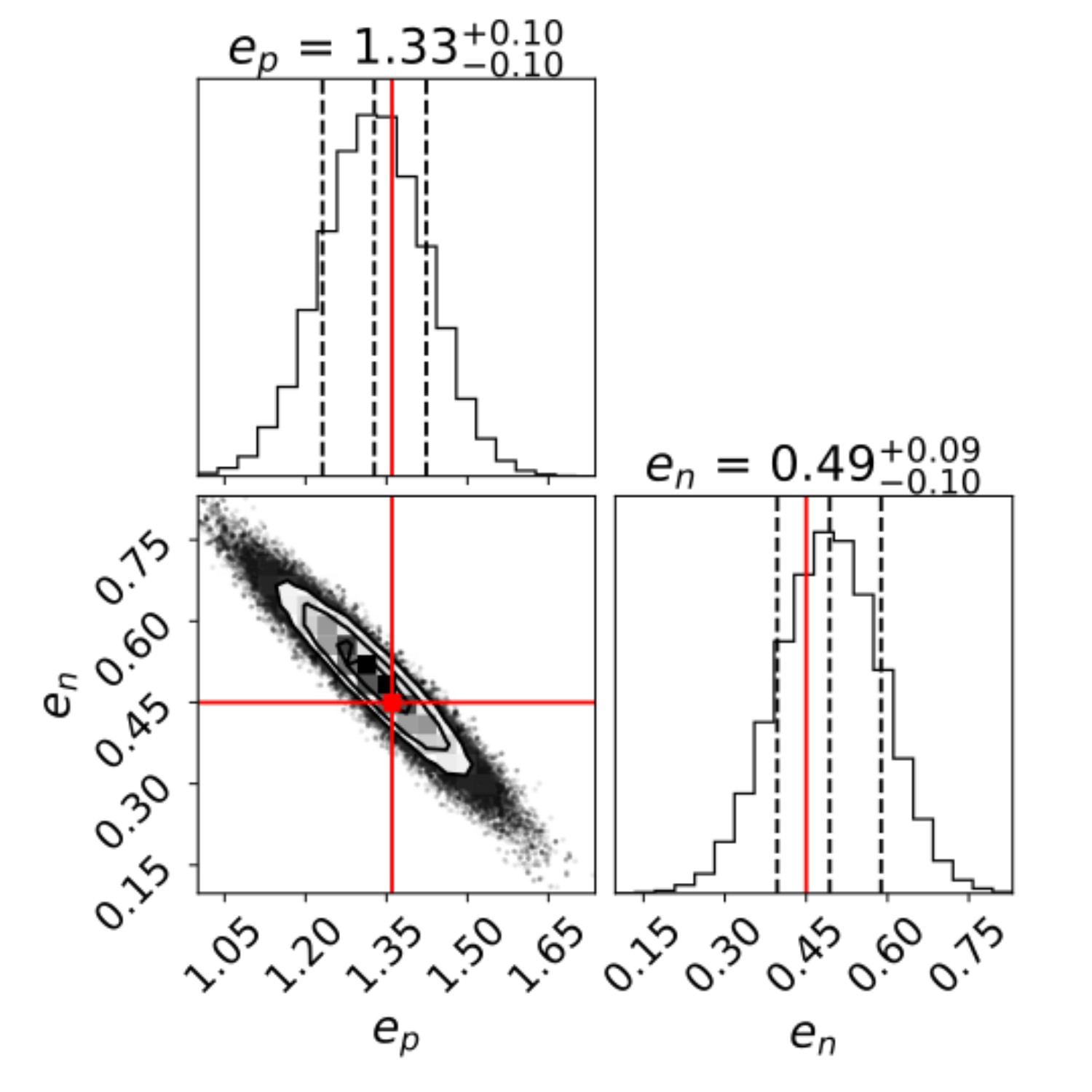} 
    \caption{Joint histogram of the effective charges for proton and neutron in $E2$ transitions, with no data truncation based on relative errors. Red lines show values from \cite{richter_sd_obs}, which were used to determine Gaussian priors. Flat priors give almost identical results.}
    \label{fig:joint_hist_eff_charges}
\end{figure}

\begin{figure}
    \includegraphics[scale=0.5]{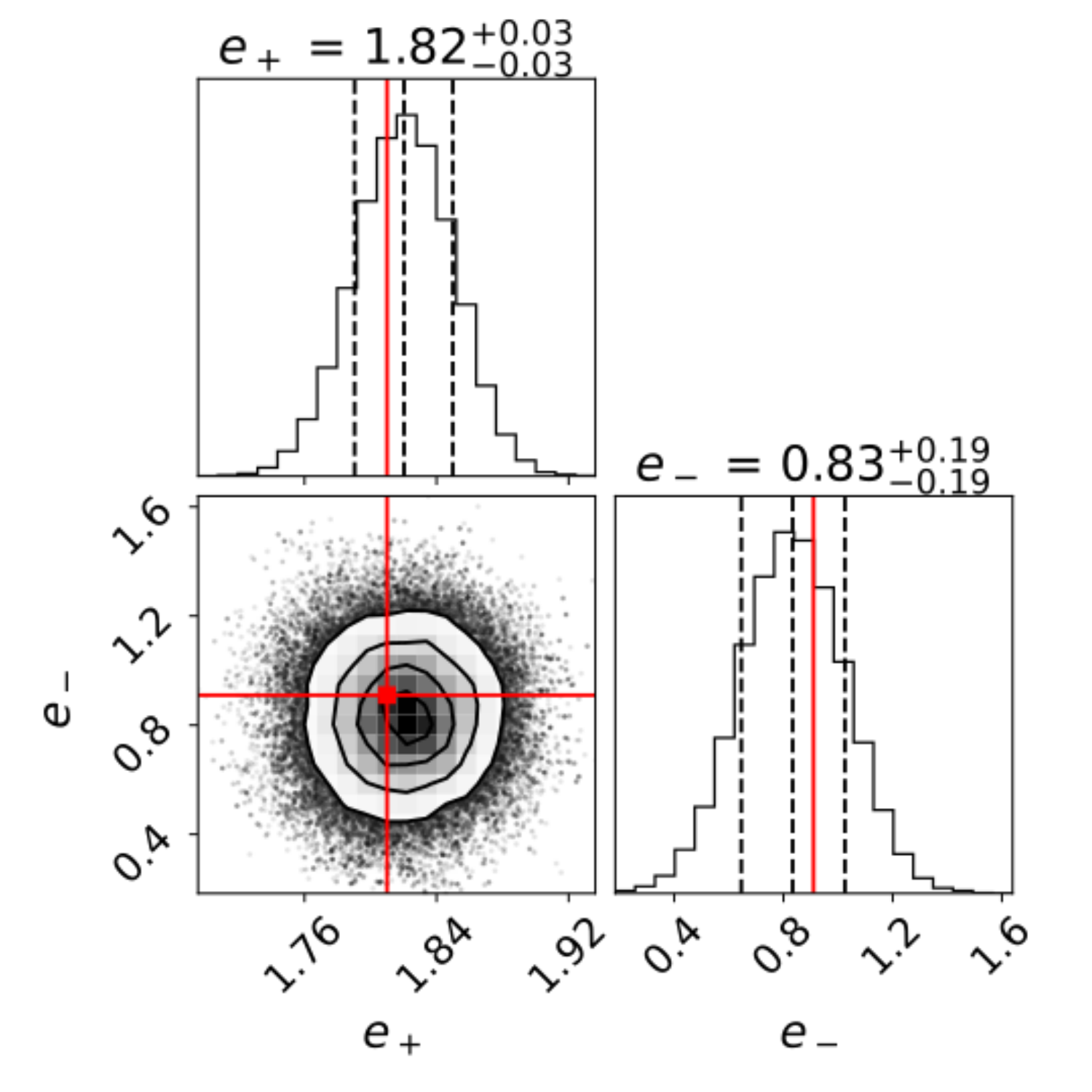} 
    \caption{Effective charges determined by $E2$ transitions, in isoscalar/isovector terms: $e_\pm = e_p \pm e_n$. We see by comparing the standard deviations that the isoscalar component $e_+$ is much more tightly constrained than the isovector component $e_-$. Note that the isospin basis fully decorrelates the parameters.}
    \label{fig:joint_hist_eff_charges_iso}
\end{figure}

Our results for $e_p, e_n$, $1.33 \pm 0.1$ and $0.49 \pm 0.1$, respectively, have central values within uncertainty of the estimates of \cite{richter_sd_obs}, although we predict larger standard deviations. 
Fig.~\ref{fig:joint_hist_eff_charges} shows the effective charges are strongly correlated, as expected.
Fig.~\ref{fig:joint_hist_eff_charges_iso} shows the same data but in the isospin basis: the isoscalar component, $e_p + e_n = 1.82 \pm 0.03$ is tightly constrained, and we see much larger uncertainty on the isovector component $e_p - e_n = 0.83 \pm 0.19$. This agrees with a recent 
\textit{ab initio} study~\cite{PhysRevC.105.034333}. 
The large uncertainty on the isovector effective charge can be explained by the $E2$ matrix elements being dominated, at least for nuclides away from shell boundaries, by the isoscalar component.

Using our results for optimal effective charges, we compute again the static theoretical uncertainty in $B(E2)$ to be 4.1 in Weisskopf units. We suggest that static theoretical uncertainties be taken with a somewhat wide confidence interval, since this value is dependent on the experimental data and the parameters $\theta$. (Additionally, since we have shown that the interaction is dominated by only 10 parameters, one could argue this uncertainty is actually smaller. For example if we compute the theoretical uncertainty again, setting the reduced chi-squared in Eq.~\ref{eq:reduced_chisq_trans} for $B(E2)$ to 1, and using $N_p=10$ rather than $66$, we get a standard deviation of 3.0 Weisskopf units.) 

\begin{figure}    \includegraphics[scale=0.4]{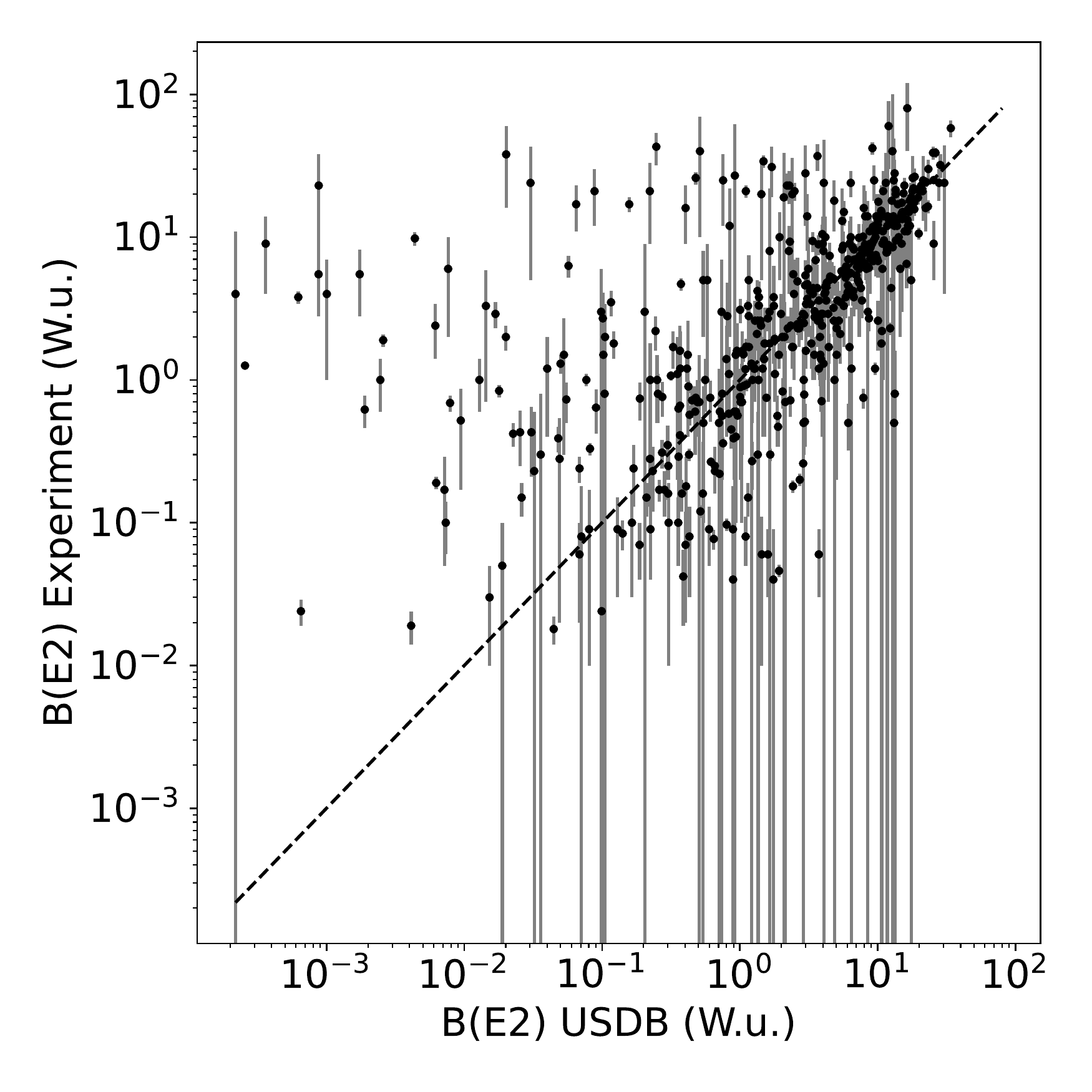} 
    \caption{Electric quadrupole transition strengths computed with the USDB interaction compared with experimental data with experimental error bars and using effective charges $(e_p,e_n)=(1.33, 0.49)$.}
    \label{fig:E2_scatter}
\end{figure}

Lastly, Fig.~\ref{fig:E2_scatter} shows a scatter plot of B($E2$) values with a number of outliers underestimated by the shell-model calculation. In general, it is more common to underestimate these transition strengths than overestimate. The reasons for this may be multiple, as experimental data is given with less precision than simulation, as well as the theory being subject to subtle cancellations. 
For these anomalous transitions, we checked whether or not they were near neutron shell closures, as intruders \cite{PhysRevC.107.034306} can mix in strongly. However, this was not generally the case, and future UQ studies may benefit from including nuclides with greater neutron excess.

\subsection{Magnetic dipole transitions and moments}

\begin{figure}
    \includegraphics[scale=0.5]{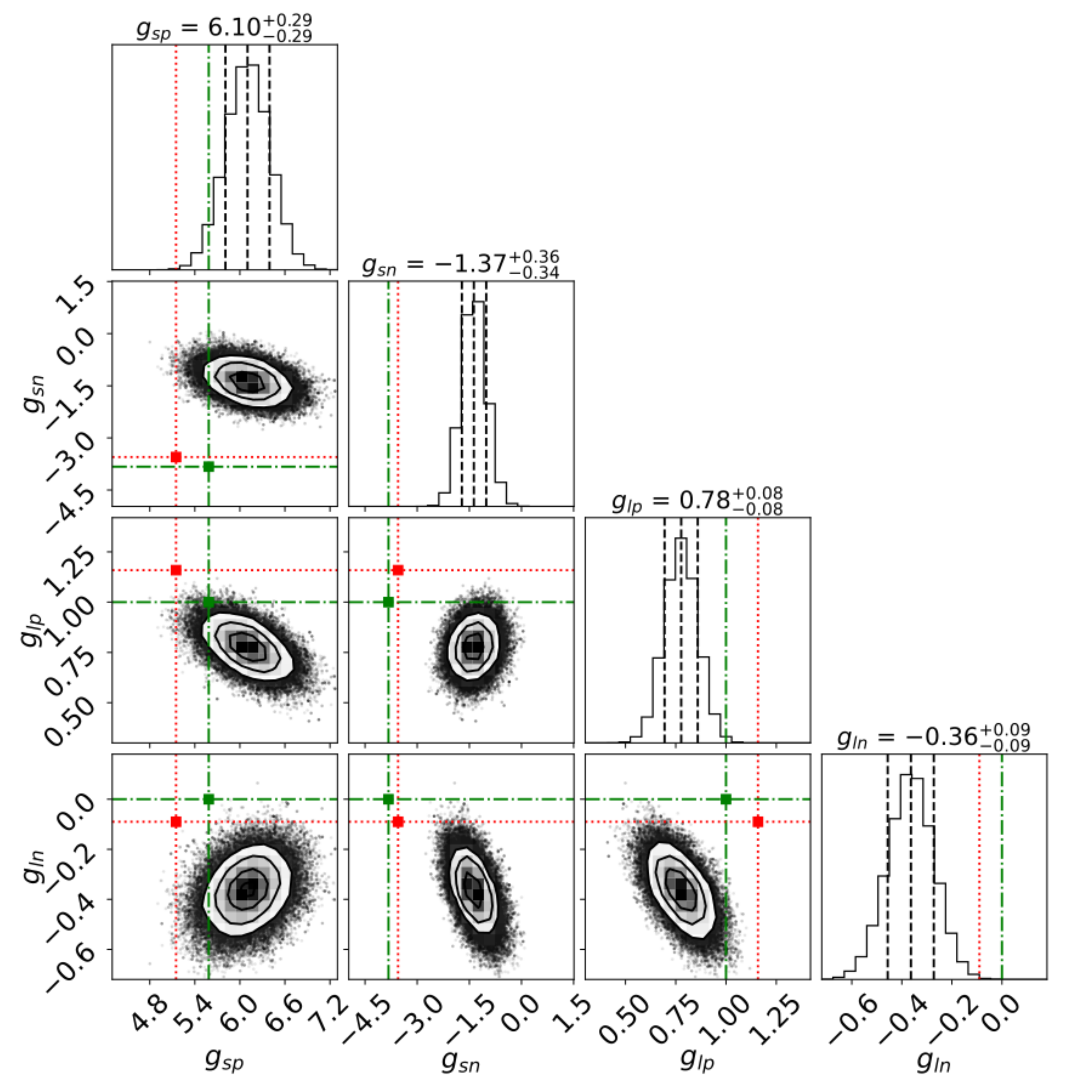} 
    \caption{Posterior distribution of the $M1$ coupling constants using a flat prior. 
    Black dashed lines show the 1-$\sigma$ band of our calculation. Green dot-dash lines show the free nucleon couplings. Red dotted lines show the results of a previous fit using the same Hamiltonian but different observations.}
    \label{fig:M1_corner_flatprior}
\end{figure}

\begin{figure}
    \includegraphics[scale=0.5]{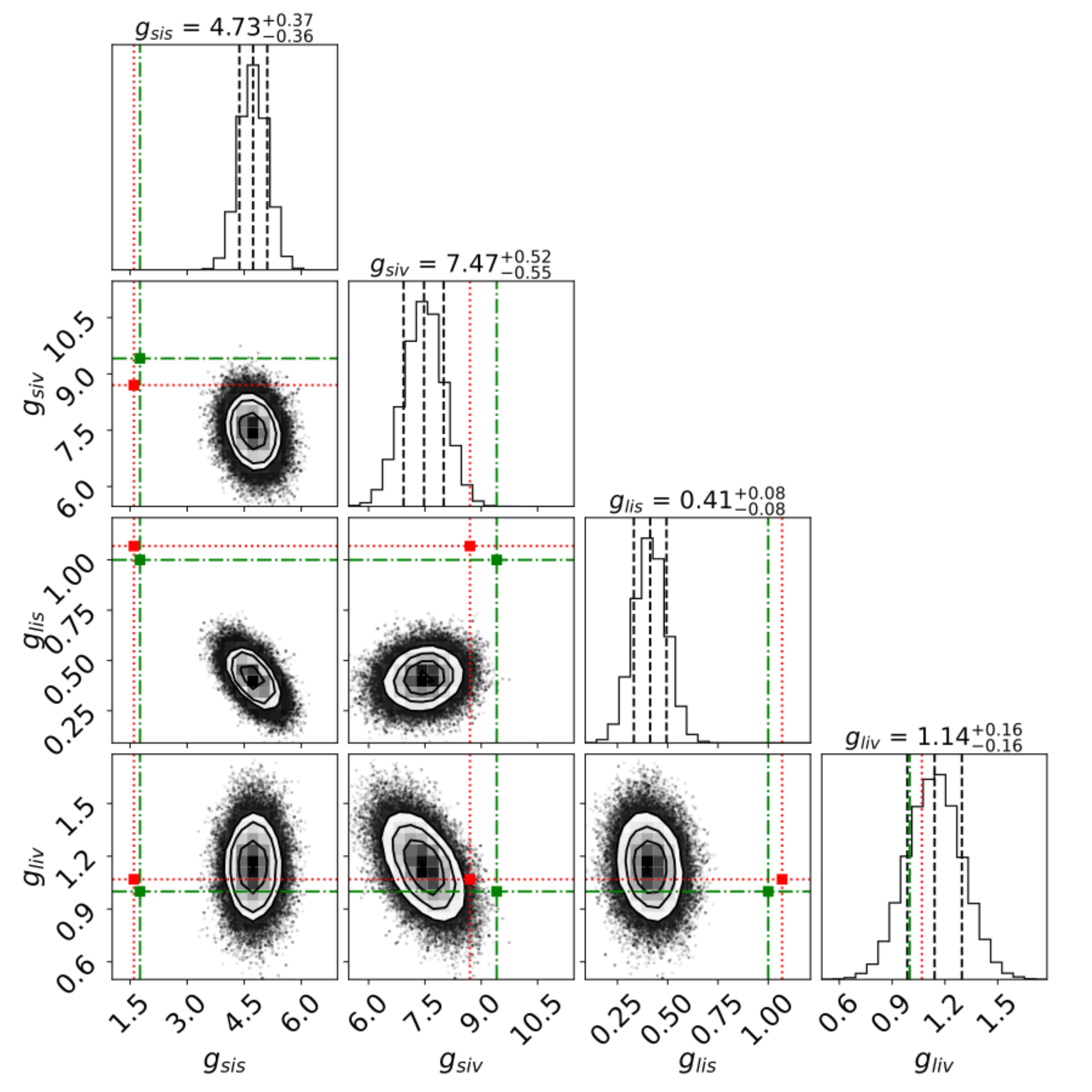} 
    \caption{Posterior distribution of the $M1$ coupling constants using a flat prior in isospin terms.
    Black dashed lines show the 1-$\sigma$ band of our calculation. Green dot-dash lines show the free nucleon couplings. Red dotted lines show the results of a previous fit using the same Hamiltonian but different observations.}
    \label{fig:M1_corner_flatprior_iso}
\end{figure}

\begin{figure}
    \includegraphics[scale=0.5]{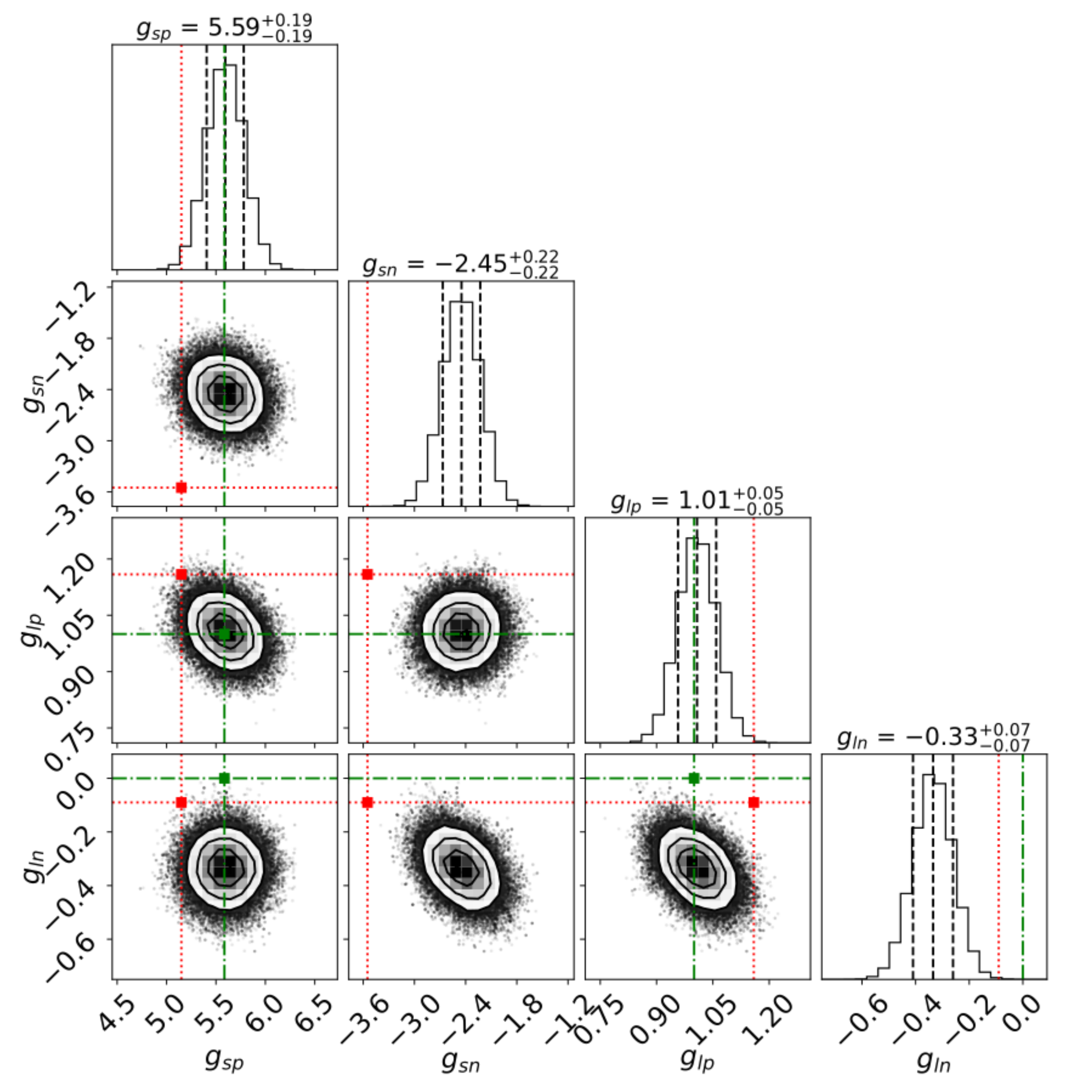} 
    \caption{Posterior distribution of the $M1$ coupling constants using a normal prior determined by optimal values listed in \cite{richter_sd_obs}.
    Black dashed lines show the 1-$\sigma$ band of our calculation. Green dot-dash lines show the free nucleon couplings. Red dotted lines show the results of a previous fit using the same Hamiltonian but different observations.}
    \label{fig:M1_corner_normalprior}
\end{figure}

\begin{figure}
    \includegraphics[scale=0.5]{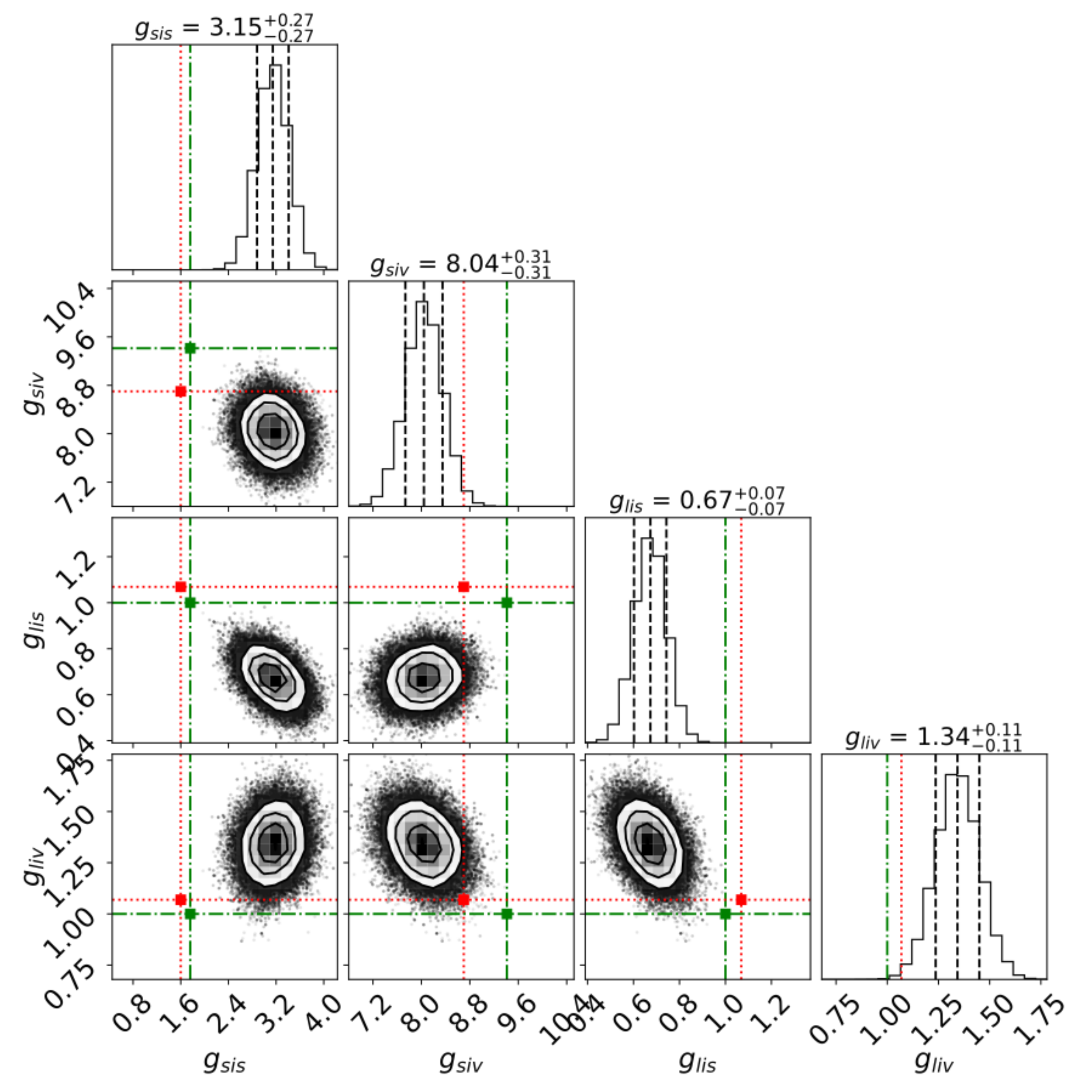} 
    \caption{Posterior distribution of the $M1$ coupling constants using a normal prior determined by optimal values listed in \cite{richter_sd_obs} in isospin terms. 
        Black dashed lines show the 1-$\sigma$ band of our calculation. Green dot-dash lines show the free nucleon couplings. Red dotted lines show the results of a previous fit using the same Hamiltonian but different observations.}
    \label{fig:M1_corner_normalprior_iso}
\end{figure}


The magnetic dipole reduced matrix element is
\begin{equation}
    M_{if}(M1) = \langle J_f || \left[ g_{sp} \vec{s}_p + g_{sn} \vec{s}_n + g_{\ell p} \vec{\ell}_p + g_{\ell n} \vec{\ell}_n \right] || J_i \rangle
    \label{eq:m1_op}
\end{equation}
where $\vec{s}_t$ is the spin operator and $\vec{\ell}_t$ is the orbital angular momentum operator, and $t=p,n$ denotes action only on the proton/neutron part of the wavefunction. The parameters of interest are the coupling constants: $\theta = (g_{sp}, g_{sn}, g_{\ell p}, g_{\ell n}) $. For the free nucleon, these coupling constants are $(5.5857,-3.8263,1,0)$, but previous work on $M1$ transitions with USDB \cite{richter_sd_obs} found optimal values of $(5.15,-3.55,1.159,-0.09)$. The slight difference in values using USDB is due to physics left out of the empirical Hamiltonian (see histograms of $M1$ matrix elements in supplemental materials), as well as optimizing for a finite number of transitions in the $sd$-shell.
Note that Eq.~\ref{eq:m1_op} excludes the $M1$ tensor operator which would allow for $\ell$-forbidden transitions. 

We also included magnetic dipole moments in this analysis, $\mu(M1) = \sqrt{\frac{4\pi}{3}} \sqrt{\frac{J}{(J+1)(2J+1)}} M(M1) $ in units of the nuclear magneton. In general, errors on moments are significantly smaller than on transition strengths.

The $M1$ transition is more challenging than the GT and $E2$ for a few reasons. Transition strengths tend to be small compared to the Weisskopf single-particle estimate, which can be due to cancellation between the four components, and/or quenching of the coupling constants similar to the quenching of $g_A$ in the Gamow-Teller. Individual matrix elements can also be highly non-Gaussian with respect to variation in the Hamiltonian. Due to the central limit theorem however, distributions of the coupling constants approach normality when summing over many experimental data points.

The results for the $B(M1)$ couplings using a flat prior are shown in Fig.~\ref{fig:M1_corner_flatprior} in proton/neutron terms and in Fig.~\ref{fig:M1_corner_flatprior_iso} in isoscalar/isovector terms. Compared to the results in \cite{richter_sd_obs}, our proton spin coupling is larger and our neutron spin coupling is significantly smaller in magnitude (by a factor of 3 roughly), but the latter term also has the largest uncertainty (25\%). Our proton orbit coupling is smaller but with the tightest uncertainty, and the neutron orbit is smaller in magnitude. The isospin picture neatly orders the uncertainty on terms: orbital isoscalar is the most constrained (as expected), then increasingly less-constrained (about $2\times$ each) we have orbital isovector, spin isoscalar, and finally spin isovector. Relative to their magnitudes, these uncertainties are roughly $20\%, 14\%, 7\%, $ and $7\%$ respectively.

We also construct informative priors from the results of \cite{richter_sd_obs}, Gaussian distributions centered at optimal values for USDB and with standard deviations at $3\times$ that of the tabulated results (The factor of 3 is chosen simply to inflate the Gaussian confidence interval from 68\% to 99.7\%) as shown here: 

\begin{equation}\label{eq:m1_prior}
    \begin{aligned}
        5.15(9) \rightarrow P(g_{sp}) &= \mathcal{N}( 5.15 , & 0.27)  \\
        -3.55(10) \rightarrow P(g_{sn}) &= \mathcal{N}( -3.55 , & 0.3 )  \\
        1.159(23) \rightarrow P(g_{\ell p}) &= \mathcal{N}( 1.159 , & 0.069 )  \\
        -0.09(26) \rightarrow P(g_{\ell n}) &= \mathcal{N}( -0.09 , & 0.78 )  \\
        \end{aligned}
\end{equation}

As expected, using informative priors gives results nearer to those previously cited. These results may act as general recommended values, but it is important to note that construction of informative priors is highly nontrivial.  

\begin{figure}
    \includegraphics[scale=0.4]{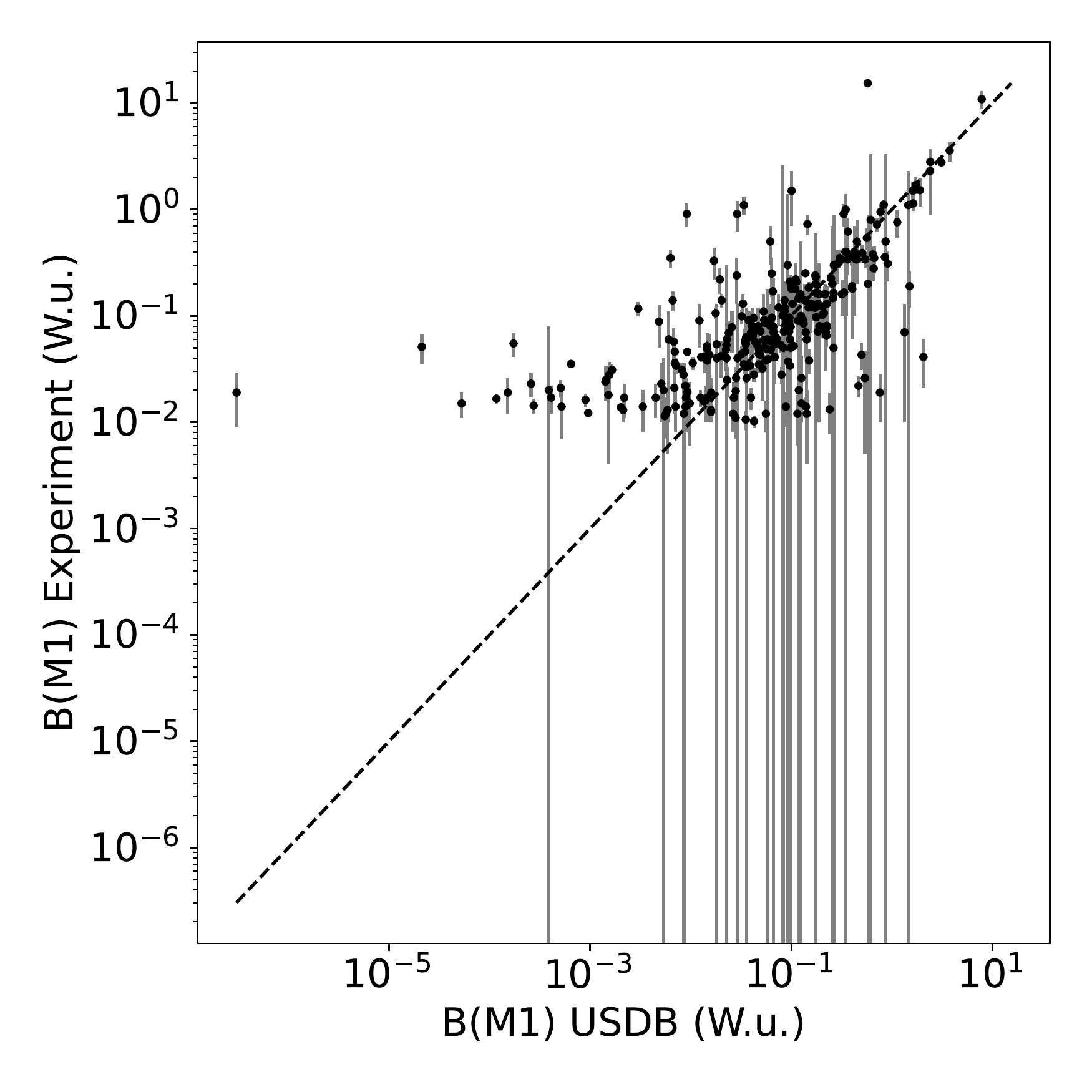} 
    \caption{Magnetic dipole transition strengths computed with the USDB interaction compared with experimental data with experimental error bars and using gyromagnetic factors $(g_{sp},g_{sn},g_{lp},g_{ln})=(5.59,-2.45,1.01,-0.33)$.}
    \label{fig:BM1_scatter}
\end{figure}

\begin{figure}
    \includegraphics[scale=0.4]{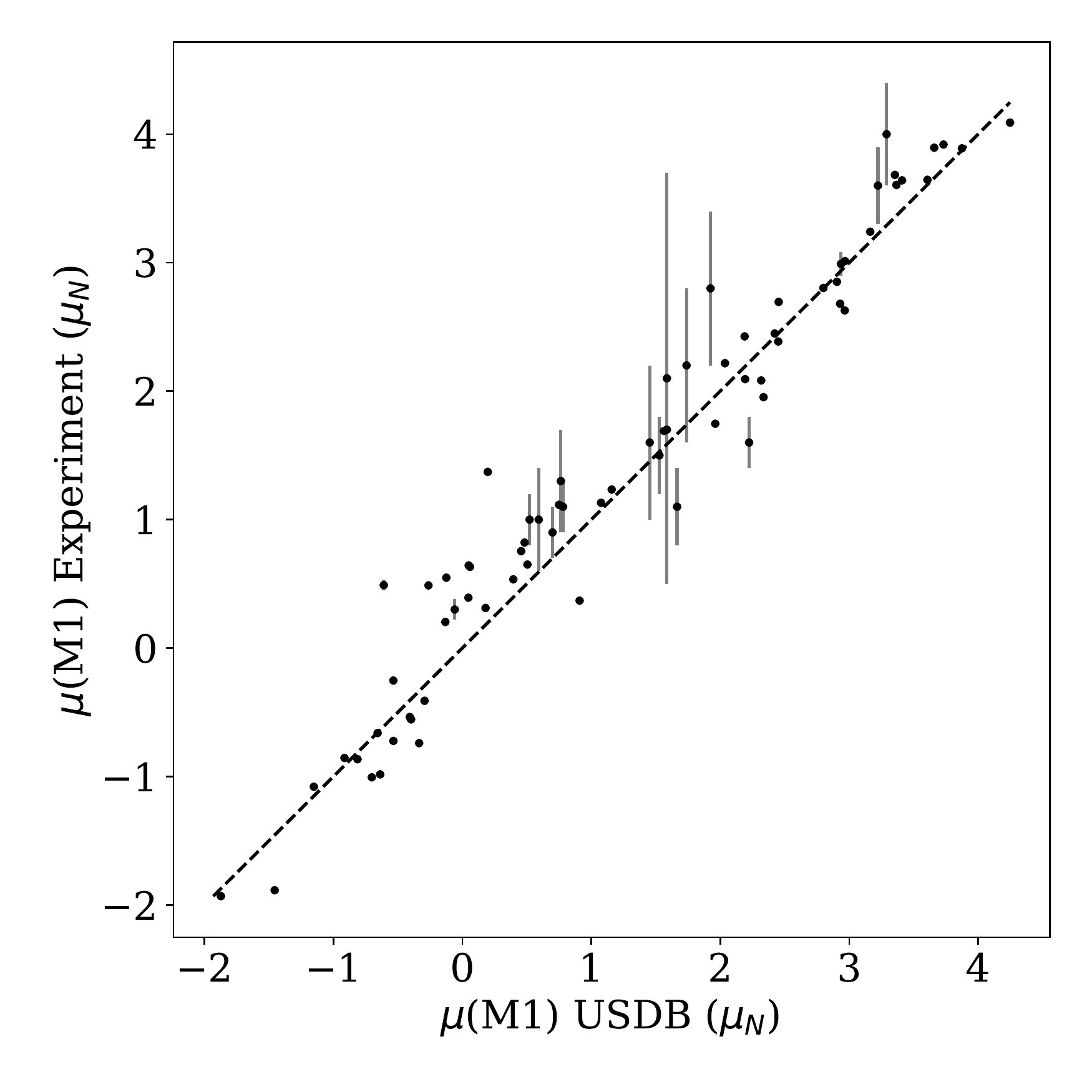} 
    \caption{Magnetic dipole moments computed with the USDB interaction compared with experimental data with experimental error bars and using gyromagnetic factors $(g_{sp},g_{sn},g_{lp},g_{ln})=(5.59,-2.45,1.01,-0.33)$.}
    \label{fig:mM1_scatter}
\end{figure}

Figures ~\ref{fig:BM1_scatter} and ~\ref{fig:mM1_scatter} show respectively a comparison of experimental $M1$ transitions and moments to the USDB calculations using the maximum-likelihood estimate of gyromagnetic factors. As with the $E2$, Fig.~\ref{fig:mM1_scatter} shows a number of outliers underestimated by the shell-model calculation, in part due to limited precision of experiment. Cancellations in the $M1$ can be troublesome, especially since the tensor contribution is omitted.

\section{Conclusions}

We have presented a method of uncertainty quantification for parameters of transition matrix elements resulting from empirical shell model calculations. Our parameters fall into two categories: those in the nuclear Hamiltonian ($\lambda$), and those in the transition operator ($\theta$). The present analysis is primarily concerned with the latter. While a fully Bayesian UQ analysis would ideally model $\lambda$ and $\theta$ together, we assign a distribution to $\lambda$ and construct a likelihood for $\theta$ based on a fixed sampling of $\lambda$. The result trades a high-resolution picture of $\lambda$ in exchange for relatively quick calculation of the $\theta$ posterior.

The work presented here fits into the larger picture of theoretical UQ in nuclear theory. While significant efforts have been underway for many years to quantify uncertainties in nuclear theory, from uncertainties in  nuclear interactions \cite{navarroperez_Bootstrapping_2014, furnstahl2015recipe, carlsson2016uncertainty, wesolowski2019exploring, PhysRevC.100.044001, perez2016uncertainty} and \textit{ab initio} calculations of light an medium nuclei \cite{binder_Initio_2014, furnstahl_Infrared_2015, wendt_Infrared_2015, ekstrom_Bayesian_2019} to mean field calculations of heavy nuclei \cite{roca-maza_Covariance_2015, schunckErrorAnalysisNuclear2015, erler_Error_2015, neufcourtQuantifiedLimitsNuclear2020} and simulations of astronomical nucleosynthesis  processes \cite{surman_Sensitivity_2014, mumpower_Impact_2015, martin_Impact_2016, mumpower_Impact_2016, mumpower_Reverse_2017, utamaRefiningMassFormulas2017}, UQ in empirical shell model has been less abundant \cite{PhysRevC.98.061301, PhysRevC.96.054316}. In particular this work is, to our knowledge, the first approach to quantifying uncertainties of transition operators.

Further research should use our results to inform more accurate UQ analyses. For instance, recent work has shown eigenvector continuation (EC) to be an extremely powerful approach to emulating eigenvalue problems. One could do away with the simplistic $P(\lambda)$ and instead define the likelihood with a EC model. That way, one may evaluate a joint posterior $P(\lambda,\theta|D)$ by MCMC and potentially get a more accurate correlation analysis.  

\section{Acknowledgements}

This material is based upon work supported by the U.S. Department of Energy, Office of Science, Office of Nuclear Physics, under Award Number  DE-FG02-03ER41272.

Computing support for this work came in part  from the Lawrence Livermore National Laboratory institutional Computing Grand Challenge program. Lawrence Livermore National Laboratory is operated by Lawrence Livermore National Security, LLC, for the U.S. Department of Energy, National Nuclear Security Administration under Contract DE-AC52-07NA27344.

JF is supported by the U.S. Department of Energy, Office of Science, Office of Nuclear Physics, under contracts DE-AC02-06CH11357, by the NUCLEI SciDAC program, and Argonne LDRD awards.

Corner plots made using the {\tt corner} library \cite{corner}.

\appendix

\section{Parameter sensitivity from energies, transitions, and sum rule operators} \label{appendix:hessian_comparisons}

In our previous study~\cite{fox_usdb} we evaluated the covariance of Hamiltonian parameters $\bm{\lambda}$ with respect to energies, i.e., Eq.~(\ref{hessian}), (\ref{eq:chisq_energy}). 
In this appendix we show some results of computing (approximate) Hessian matrices of $\bm{\lambda}$ with respect to energy, which was used for our sensitivity analysis in this paper, and again using $B(\text{GT})$ values instead of energies.
The results of using other observables, $B(M1)$, $B(E2)$ and sum rule operators, are plotted and included in supplemental materials.
The (non-energy-weighted) sum-rule operator for transition $\hat{O}$ is $\hat{O}^\dagger \hat{O}$, and is so called because the expectation value implicitly counts up transition contributions over final states: $\langle \hat{O}^\dagger \hat{O} \rangle_i \propto \sum_f \langle i | \hat{O}^\dagger |f \rangle \langle f | \hat{O} | i \rangle$.

We compute all Hessians by the approximation
$H_\lambda \approx  J_{\Omega(\lambda)}^T C_\Omega^{-1} J_{\Omega(\lambda)} $ where $\Omega$ is the observable (energy, 
Gamow-Teller strengths, sum rules) and 
$\left[ J_{\Omega(\lambda)} \right]_{i \alpha } = \partial \Omega^\text{th}_\alpha (\bm{\lambda}) / \partial \lambda_i$, 
and $C_\Omega$ is the diagonal matrix of errors.

\begin{figure}
    \includegraphics[scale=0.4,clip]{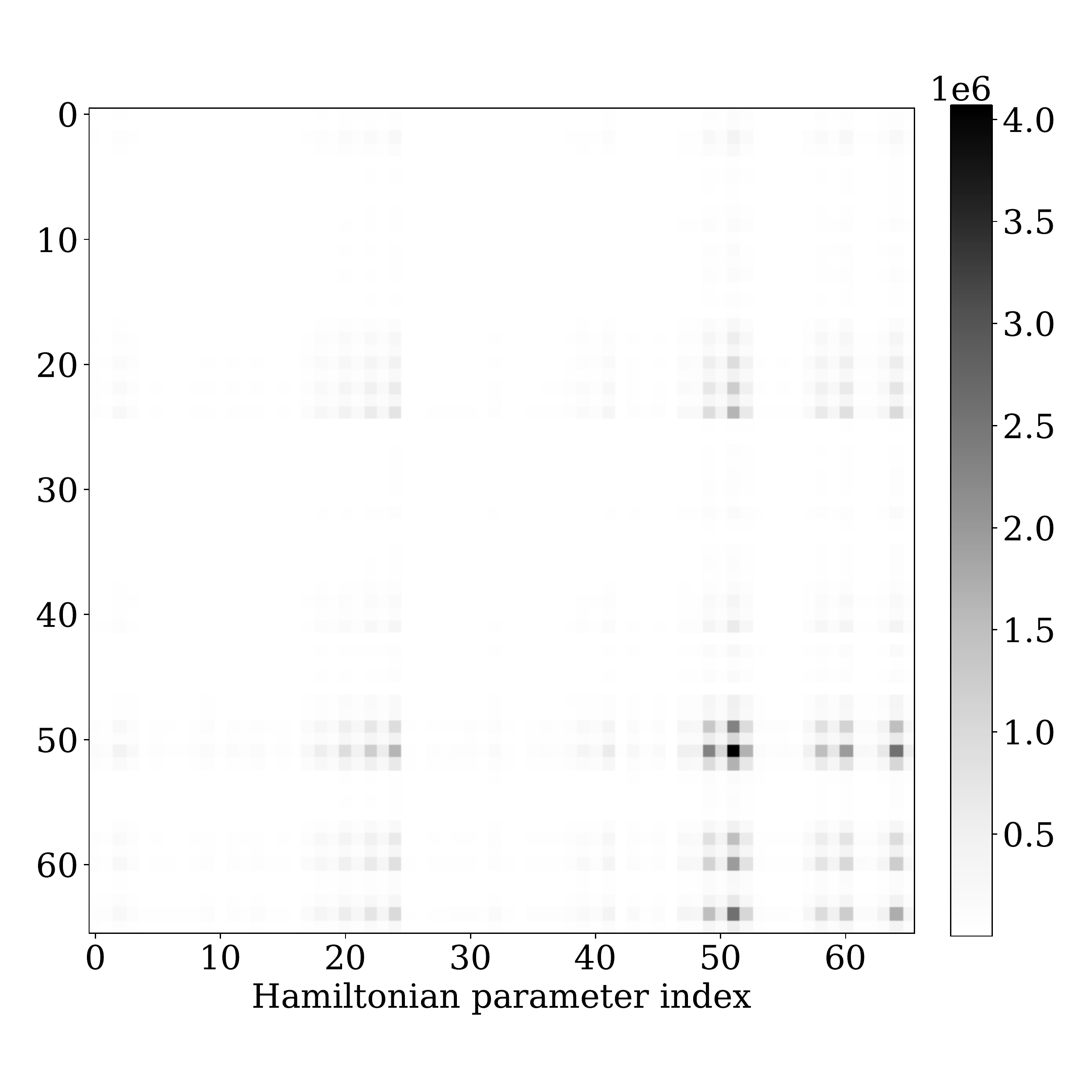} 
    \caption{$A_{\lambda(E)}$, approximate Hessian matrix for Hamiltonian parameters $\bm{\lambda}$, computed from energies. The darkest cells indicate the most important matrix elements for energies.}
    \label{fig:hessian_E}
\end{figure}

The Hessian for energies, Eq.~(\ref{hessian}), (\ref{eq:chisq_energy}), is plotted in Figure \ref{fig:hessian_E}. The energy is dominated by two TBMEs (\#51, and 60 here): isovector pairs $(0d_{5/2})(0d_{5/2})$ with $J=4$, and isovector pairs $(0d_{5/2})(1s_{1/2})$ with $J=3$.
These are matrix elements between normalized 
two-body states $| a b; J T \rangle $ with nucleons in orbitals $a$
and $b$ coupled up to total angular momentum $J$ and total isospin $T$;
the commonplace notation for the shell model is $V_{JT}(ab,cd)$.
The GT is dominated by the same two matrix elements, and $E2$ and $M1$ are similar as well (see supplemental materials).

\begin{figure}
    \includegraphics[scale=0.4,clip]{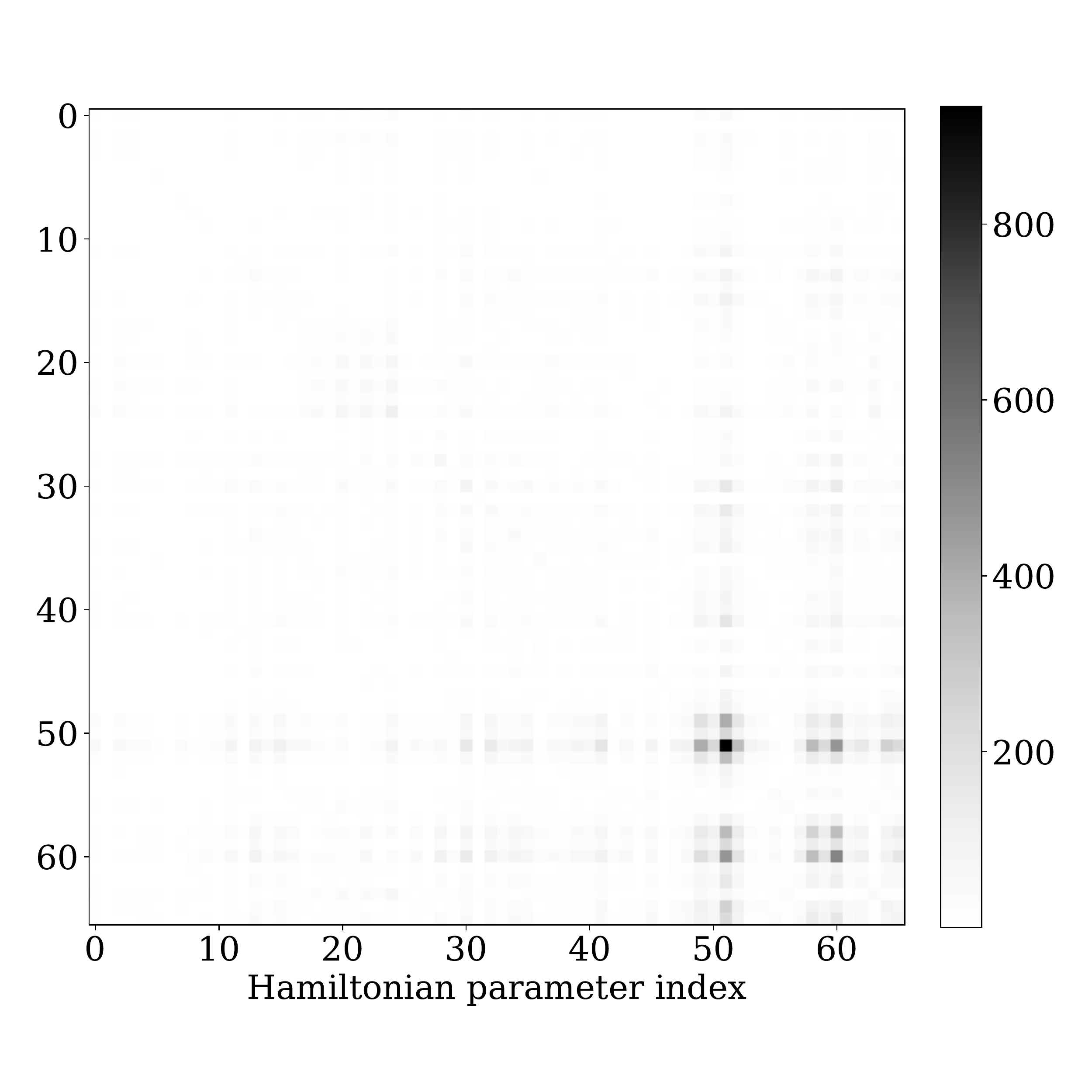} 
    \caption{$A_{\lambda(B)}$, approximate Hessian matrix for Hamiltonian parameters $\bm{\lambda}$, computed from Gamow-Teller transition strengths. The darkest cells indicate the most important matrix elements for $B(\text{GT})$. Note the similarity to $A_{\lambda(E)}$, shown in Fig.~\ref{fig:hessian_E}.}
    \label{fig:hessian_GT}
\end{figure}


\bibliography{johnsonmaster,fox_uq,navarroperez}

\begin{thebibliography}{72}%
\makeatletter
\providecommand \@ifxundefined [1]{%
 \@ifx{#1\undefined}
}%
\providecommand \@ifnum [1]{%
 \ifnum #1\expandafter \@firstoftwo
 \else \expandafter \@secondoftwo
 \fi
}%
\providecommand \@ifx [1]{%
 \ifx #1\expandafter \@firstoftwo
 \else \expandafter \@secondoftwo
 \fi
}%
\providecommand \natexlab [1]{#1}%
\providecommand \enquote  [1]{``#1''}%
\providecommand \bibnamefont  [1]{#1}%
\providecommand \bibfnamefont [1]{#1}%
\providecommand \citenamefont [1]{#1}%
\providecommand \href@noop [0]{\@secondoftwo}%
\providecommand \href [0]{\begingroup \@sanitize@url \@href}%
\providecommand \@href[1]{\@@startlink{#1}\@@href}%
\providecommand \@@href[1]{\endgroup#1\@@endlink}%
\providecommand \@sanitize@url [0]{\catcode `\\12\catcode `\$12\catcode
  `\&12\catcode `\#12\catcode `\^12\catcode `\_12\catcode `\%12\relax}%
\providecommand \@@startlink[1]{}%
\providecommand \@@endlink[0]{}%
\providecommand \url  [0]{\begingroup\@sanitize@url \@url }%
\providecommand \@url [1]{\endgroup\@href {#1}{\urlprefix }}%
\providecommand \urlprefix  [0]{URL }%
\providecommand \Eprint [0]{\href }%
\providecommand \doibase [0]{https://doi.org/}%
\providecommand \selectlanguage [0]{\@gobble}%
\providecommand \bibinfo  [0]{\@secondoftwo}%
\providecommand \bibfield  [0]{\@secondoftwo}%
\providecommand \translation [1]{[#1]}%
\providecommand \BibitemOpen [0]{}%
\providecommand \bibitemStop [0]{}%
\providecommand \bibitemNoStop [0]{.\EOS\space}%
\providecommand \EOS [0]{\spacefactor3000\relax}%
\providecommand \BibitemShut  [1]{\csname bibitem#1\endcsname}%
\let\auto@bib@innerbib\@empty
\bibitem [{\citenamefont {Dobaczewski}\ \emph {et~al.}(2014)\citenamefont
  {Dobaczewski}, \citenamefont {Nazarewicz},\ and\ \citenamefont
  {Reinhard}}]{dobaczewski2014error}%
  \BibitemOpen
  \bibfield  {author} {\bibinfo {author} {\bibfnamefont {J.}~\bibnamefont
  {Dobaczewski}}, \bibinfo {author} {\bibfnamefont {W.}~\bibnamefont
  {Nazarewicz}},\ and\ \bibinfo {author} {\bibfnamefont {P.}~\bibnamefont
  {Reinhard}},\ }\bibfield  {title} {\bibinfo {title} {Error estimates of
  theoretical models: a guide},\ }\href@noop {} {\bibfield  {journal} {\bibinfo
   {journal} {Journal of Physics G: Nuclear and Particle Physics}\ }\textbf
  {\bibinfo {volume} {41}},\ \bibinfo {pages} {074001} (\bibinfo {year}
  {2014})}\BibitemShut {NoStop}%
\bibitem [{\citenamefont {{Navarro P{\'e}rez}}\ \emph
  {et~al.}(2015)\citenamefont {{Navarro P{\'e}rez}}, \citenamefont {{Amaro}},\
  and\ \citenamefont {{Ruiz Arriola}}}]{navarro_perez_error_analysis_2015}%
  \BibitemOpen
  \bibfield  {author} {\bibinfo {author} {\bibfnamefont {R.}~\bibnamefont
  {{Navarro P{\'e}rez}}}, \bibinfo {author} {\bibfnamefont {J.~E.}\
  \bibnamefont {{Amaro}}},\ and\ \bibinfo {author} {\bibfnamefont
  {E.}~\bibnamefont {{Ruiz Arriola}}},\ }\bibfield  {title} {\bibinfo {title}
  {{Error analysis of nuclear forces and effective interactions}},\ }\href
  {https://doi.org/10.1088/0954-3899/42/3/034013} {\bibfield  {journal}
  {\bibinfo  {journal} {Journal of Physics G Nuclear Physics}\ }\textbf
  {\bibinfo {volume} {42}},\ \bibinfo {eid} {034013} (\bibinfo {year}
  {2015})},\ \Eprint {https://arxiv.org/abs/1406.0625} {arXiv:1406.0625
  [nucl-th]} \BibitemShut {NoStop}%
\bibitem [{\citenamefont {Xu}\ \emph {et~al.}(2021)\citenamefont {Xu},
  \citenamefont {Zhang},\ and\ \citenamefont {Li}}]{xu2021bayesian}%
  \BibitemOpen
  \bibfield  {author} {\bibinfo {author} {\bibfnamefont {J.}~\bibnamefont
  {Xu}}, \bibinfo {author} {\bibfnamefont {Z.}~\bibnamefont {Zhang}},\ and\
  \bibinfo {author} {\bibfnamefont {B.-A.}\ \bibnamefont {Li}},\ }\bibfield
  {title} {\bibinfo {title} {Bayesian uncertainty quantification for nuclear
  matter incompressibility},\ }\href@noop {} {\bibfield  {journal} {\bibinfo
  {journal} {Physical Review C}\ }\textbf {\bibinfo {volume} {104}},\ \bibinfo
  {pages} {054324} (\bibinfo {year} {2021})}\BibitemShut {NoStop}%
\bibitem [{\citenamefont {Lovell}\ \emph {et~al.}(2020)\citenamefont {Lovell},
  \citenamefont {Nunes}, \citenamefont {Catacora-Rios},\ and\ \citenamefont
  {King}}]{Lovell_2020}%
  \BibitemOpen
  \bibfield  {author} {\bibinfo {author} {\bibfnamefont {A.~E.}\ \bibnamefont
  {Lovell}}, \bibinfo {author} {\bibfnamefont {F.~M.}\ \bibnamefont {Nunes}},
  \bibinfo {author} {\bibfnamefont {M.}~\bibnamefont {Catacora-Rios}},\ and\
  \bibinfo {author} {\bibfnamefont {G.~B.}\ \bibnamefont {King}},\ }\bibfield
  {title} {\bibinfo {title} {Recent advances in the quantification of
  uncertainties in reaction theory},\ }\href
  {https://doi.org/10.1088/1361-6471/abba72} {\bibfield  {journal} {\bibinfo
  {journal} {Journal of Physics G: Nuclear and Particle Physics}\ }\textbf
  {\bibinfo {volume} {48}},\ \bibinfo {pages} {014001} (\bibinfo {year}
  {2020})}\BibitemShut {NoStop}%
\bibitem [{\citenamefont {Piekarewicz}\ \emph {et~al.}(2015)\citenamefont
  {Piekarewicz}, \citenamefont {Chen},\ and\ \citenamefont
  {Fattoyev}}]{piekarewicz2015information}%
  \BibitemOpen
  \bibfield  {author} {\bibinfo {author} {\bibfnamefont {J.}~\bibnamefont
  {Piekarewicz}}, \bibinfo {author} {\bibfnamefont {W.-C.}\ \bibnamefont
  {Chen}},\ and\ \bibinfo {author} {\bibfnamefont {F.}~\bibnamefont
  {Fattoyev}},\ }\bibfield  {title} {\bibinfo {title} {Information and
  statistics: a new paradigm in theoretical nuclear physics},\ }\href@noop {}
  {\bibfield  {journal} {\bibinfo  {journal} {Journal of Physics G: nuclear and
  particle physics}\ }\textbf {\bibinfo {volume} {42}},\ \bibinfo {pages}
  {034018} (\bibinfo {year} {2015})}\BibitemShut {NoStop}%
\bibitem [{\citenamefont {Whitehead}\ \emph {et~al.}(2021)\citenamefont
  {Whitehead}, \citenamefont {Poxon-Pearson}, \citenamefont {Nunes},\ and\
  \citenamefont {Potel}}]{whitehead2021prediction}%
  \BibitemOpen
  \bibfield  {author} {\bibinfo {author} {\bibfnamefont {T.}~\bibnamefont
  {Whitehead}}, \bibinfo {author} {\bibfnamefont {T.}~\bibnamefont
  {Poxon-Pearson}}, \bibinfo {author} {\bibfnamefont {F.}~\bibnamefont
  {Nunes}},\ and\ \bibinfo {author} {\bibfnamefont {G.}~\bibnamefont {Potel}},\
  }\bibfield  {title} {\bibinfo {title} {Prediction of (p, n) charge-exchange
  reactions with uncertainty quantification},\ }\href@noop {} {\bibfield
  {journal} {\bibinfo  {journal} {arXiv preprint arXiv:2112.14256}\ } (\bibinfo
  {year} {2021})}\BibitemShut {NoStop}%
\bibitem [{\citenamefont {Drischler}\ \emph {et~al.}(2020)\citenamefont
  {Drischler}, \citenamefont {Melendez}, \citenamefont {Furnstahl},\ and\
  \citenamefont {Phillips}}]{drischler2020quantifying}%
  \BibitemOpen
  \bibfield  {author} {\bibinfo {author} {\bibfnamefont {C.}~\bibnamefont
  {Drischler}}, \bibinfo {author} {\bibfnamefont {J.}~\bibnamefont {Melendez}},
  \bibinfo {author} {\bibfnamefont {R.}~\bibnamefont {Furnstahl}},\ and\
  \bibinfo {author} {\bibfnamefont {D.}~\bibnamefont {Phillips}},\ }\bibfield
  {title} {\bibinfo {title} {Quantifying uncertainties and correlations in the
  nuclear-matter equation of state},\ }\href@noop {} {\bibfield  {journal}
  {\bibinfo  {journal} {Physical Review C}\ }\textbf {\bibinfo {volume}
  {102}},\ \bibinfo {pages} {054315} (\bibinfo {year} {2020})}\BibitemShut
  {NoStop}%
\bibitem [{\citenamefont {P{\'e}rez}\ and\ \citenamefont
  {Arriola}(2020)}]{perez2020uncertainty}%
  \BibitemOpen
  \bibfield  {author} {\bibinfo {author} {\bibfnamefont {R.~N.}\ \bibnamefont
  {P{\'e}rez}}\ and\ \bibinfo {author} {\bibfnamefont {E.~R.}\ \bibnamefont
  {Arriola}},\ }\bibfield  {title} {\bibinfo {title} {Uncertainty
  quantification and falsification of chiral nuclear potentials},\ }\href@noop
  {} {\bibfield  {journal} {\bibinfo  {journal} {The European Physical Journal
  A}\ }\textbf {\bibinfo {volume} {56}},\ \bibinfo {pages} {1} (\bibinfo {year}
  {2020})}\BibitemShut {NoStop}%
\bibitem [{\citenamefont {Jaganathen}\ \emph {et~al.}(2017)\citenamefont
  {Jaganathen}, \citenamefont {Betan}, \citenamefont {Michel}, \citenamefont
  {Nazarewicz},\ and\ \citenamefont {P\l{}oszajczak}}]{PhysRevC.96.054316}%
  \BibitemOpen
  \bibfield  {author} {\bibinfo {author} {\bibfnamefont {Y.}~\bibnamefont
  {Jaganathen}}, \bibinfo {author} {\bibfnamefont {R.~M.~I.}\ \bibnamefont
  {Betan}}, \bibinfo {author} {\bibfnamefont {N.}~\bibnamefont {Michel}},
  \bibinfo {author} {\bibfnamefont {W.}~\bibnamefont {Nazarewicz}},\ and\
  \bibinfo {author} {\bibfnamefont {M.}~\bibnamefont {P\l{}oszajczak}},\
  }\bibfield  {title} {\bibinfo {title} {Quantified gamow shell model
  interaction for $psd$-shell nuclei},\ }\href
  {https://doi.org/10.1103/PhysRevC.96.054316} {\bibfield  {journal} {\bibinfo
  {journal} {Phys. Rev. C}\ }\textbf {\bibinfo {volume} {96}},\ \bibinfo
  {pages} {054316} (\bibinfo {year} {2017})}\BibitemShut {NoStop}%
\bibitem [{\citenamefont {Yoshida}\ \emph {et~al.}(2018)\citenamefont
  {Yoshida}, \citenamefont {Shimizu}, \citenamefont {Togashi},\ and\
  \citenamefont {Otsuka}}]{PhysRevC.98.061301}%
  \BibitemOpen
  \bibfield  {author} {\bibinfo {author} {\bibfnamefont {S.}~\bibnamefont
  {Yoshida}}, \bibinfo {author} {\bibfnamefont {N.}~\bibnamefont {Shimizu}},
  \bibinfo {author} {\bibfnamefont {T.}~\bibnamefont {Togashi}},\ and\ \bibinfo
  {author} {\bibfnamefont {T.}~\bibnamefont {Otsuka}},\ }\bibfield  {title}
  {\bibinfo {title} {Uncertainty quantification in the nuclear shell model},\
  }\href {https://doi.org/10.1103/PhysRevC.98.061301} {\bibfield  {journal}
  {\bibinfo  {journal} {Phys. Rev. C}\ }\textbf {\bibinfo {volume} {98}},\
  \bibinfo {pages} {061301} (\bibinfo {year} {2018})}\BibitemShut {NoStop}%
\bibitem [{\citenamefont {Balantekin}\ \emph {et~al.}(2014)\citenamefont
  {Balantekin}, \citenamefont {Carlson}, \citenamefont {Dean}, \citenamefont
  {Fuller}, \citenamefont {Furnstahl}, \citenamefont {{Hjorth-Jensen}},
  \citenamefont {Janssens}, \citenamefont {Li}, \citenamefont {Nazarewicz},
  \citenamefont {Nunes}, \citenamefont {Ormand}, \citenamefont {Reddy},\ and\
  \citenamefont {Sherrill}}]{balantekinNuclearTheoryScience2014}%
  \BibitemOpen
  \bibfield  {author} {\bibinfo {author} {\bibfnamefont {A.~B.}\ \bibnamefont
  {Balantekin}}, \bibinfo {author} {\bibfnamefont {J.}~\bibnamefont {Carlson}},
  \bibinfo {author} {\bibfnamefont {D.~J.}\ \bibnamefont {Dean}}, \bibinfo
  {author} {\bibfnamefont {G.~M.}\ \bibnamefont {Fuller}}, \bibinfo {author}
  {\bibfnamefont {R.~J.}\ \bibnamefont {Furnstahl}}, \bibinfo {author}
  {\bibfnamefont {M.}~\bibnamefont {{Hjorth-Jensen}}}, \bibinfo {author}
  {\bibfnamefont {R.~V.~F.}\ \bibnamefont {Janssens}}, \bibinfo {author}
  {\bibfnamefont {B.-A.}\ \bibnamefont {Li}}, \bibinfo {author} {\bibfnamefont
  {W.}~\bibnamefont {Nazarewicz}}, \bibinfo {author} {\bibfnamefont {F.~M.}\
  \bibnamefont {Nunes}}, \bibinfo {author} {\bibfnamefont {W.~E.}\ \bibnamefont
  {Ormand}}, \bibinfo {author} {\bibfnamefont {S.}~\bibnamefont {Reddy}},\ and\
  \bibinfo {author} {\bibfnamefont {B.~M.}\ \bibnamefont {Sherrill}},\
  }\bibfield  {title} {\bibinfo {title} {Nuclear theory and science of the
  facility for rare isotope beams},\ }\href
  {https://doi.org/10.1142/S0217732314300109} {\bibfield  {journal} {\bibinfo
  {journal} {Mod. Phys. Lett. A}\ }\textbf {\bibinfo {volume} {29}},\ \bibinfo
  {pages} {1430010} (\bibinfo {year} {2014})}\BibitemShut {NoStop}%
\bibitem [{\citenamefont {Acciarri}\ \emph {et~al.}(2016)\citenamefont
  {Acciarri}, \citenamefont {Bansal}, \citenamefont {Friedland}, \citenamefont
  {Djurcic}, \citenamefont {Rakotondravohitra}, \citenamefont {Xin},
  \citenamefont {Mazzucato}, \citenamefont {Densham}, \citenamefont {Calvo},
  \citenamefont {Li} \emph {et~al.}}]{acciarri2016long}%
  \BibitemOpen
  \bibfield  {author} {\bibinfo {author} {\bibfnamefont {R.}~\bibnamefont
  {Acciarri}}, \bibinfo {author} {\bibfnamefont {S.}~\bibnamefont {Bansal}},
  \bibinfo {author} {\bibfnamefont {A.}~\bibnamefont {Friedland}}, \bibinfo
  {author} {\bibfnamefont {Z.}~\bibnamefont {Djurcic}}, \bibinfo {author}
  {\bibfnamefont {L.}~\bibnamefont {Rakotondravohitra}}, \bibinfo {author}
  {\bibfnamefont {T.}~\bibnamefont {Xin}}, \bibinfo {author} {\bibfnamefont
  {E.}~\bibnamefont {Mazzucato}}, \bibinfo {author} {\bibfnamefont
  {C.}~\bibnamefont {Densham}}, \bibinfo {author} {\bibfnamefont
  {E.}~\bibnamefont {Calvo}}, \bibinfo {author} {\bibfnamefont
  {S.}~\bibnamefont {Li}}, \emph {et~al.},\ }\href@noop {} {\emph {\bibinfo
  {title} {Long-Baseline Neutrino Facility (LBNF) and Deep Underground Neutrino
  Experiment (DUNE): Volume 1: The LBNF and DUNE Projects}}},\ \bibinfo {type}
  {Tech. Rep.}\ (\bibinfo {year} {2016})\BibitemShut {NoStop}%
\bibitem [{\citenamefont {Undagoitia}\ and\ \citenamefont
  {Rauch}(2015)}]{undagoitia2015dark}%
  \BibitemOpen
  \bibfield  {author} {\bibinfo {author} {\bibfnamefont {T.~M.}\ \bibnamefont
  {Undagoitia}}\ and\ \bibinfo {author} {\bibfnamefont {L.}~\bibnamefont
  {Rauch}},\ }\bibfield  {title} {\bibinfo {title} {Dark matter
  direct-detection experiments},\ }\href@noop {} {\bibfield  {journal}
  {\bibinfo  {journal} {Journal of Physics G: Nuclear and Particle Physics}\
  }\textbf {\bibinfo {volume} {43}},\ \bibinfo {pages} {013001} (\bibinfo
  {year} {2015})}\BibitemShut {NoStop}%
\bibitem [{\citenamefont {Engel}\ and\ \citenamefont
  {Men{\'e}ndez}(2017)}]{engel2017status}%
  \BibitemOpen
  \bibfield  {author} {\bibinfo {author} {\bibfnamefont {J.}~\bibnamefont
  {Engel}}\ and\ \bibinfo {author} {\bibfnamefont {J.}~\bibnamefont
  {Men{\'e}ndez}},\ }\bibfield  {title} {\bibinfo {title} {Status and future of
  nuclear matrix elements for neutrinoless double-beta decay: a review},\
  }\href@noop {} {\bibfield  {journal} {\bibinfo  {journal} {Reports on
  Progress in Physics}\ }\textbf {\bibinfo {volume} {80}},\ \bibinfo {pages}
  {046301} (\bibinfo {year} {2017})}\BibitemShut {NoStop}%
\bibitem [{\citenamefont {Dolinski}\ \emph {et~al.}(2019)\citenamefont
  {Dolinski}, \citenamefont {Poon},\ and\ \citenamefont
  {Rodejohann}}]{dolinski2019neutrinoless}%
  \BibitemOpen
  \bibfield  {author} {\bibinfo {author} {\bibfnamefont {M.~J.}\ \bibnamefont
  {Dolinski}}, \bibinfo {author} {\bibfnamefont {A.~W.}\ \bibnamefont {Poon}},\
  and\ \bibinfo {author} {\bibfnamefont {W.}~\bibnamefont {Rodejohann}},\
  }\bibfield  {title} {\bibinfo {title} {Neutrinoless double-beta decay: Status
  and prospects},\ }\href@noop {} {\bibfield  {journal} {\bibinfo  {journal}
  {Annual Review of Nuclear and Particle Science}\ }\textbf {\bibinfo {volume}
  {69}} (\bibinfo {year} {2019})}\BibitemShut {NoStop}%
\bibitem [{\citenamefont {Kennedy}\ and\ \citenamefont
  {O'Hagan}(2001)}]{kennedy2001bayesian}%
  \BibitemOpen
  \bibfield  {author} {\bibinfo {author} {\bibfnamefont {M.~C.}\ \bibnamefont
  {Kennedy}}\ and\ \bibinfo {author} {\bibfnamefont {A.}~\bibnamefont
  {O'Hagan}},\ }\bibfield  {title} {\bibinfo {title} {Bayesian calibration of
  computer models},\ }\href
  {https://doi.org/https://doi.org/10.1111/1467-9868.00294} {\bibfield
  {journal} {\bibinfo  {journal} {Journal of the Royal Statistical Society:
  Series B (Statistical Methodology)}\ }\textbf {\bibinfo {volume} {63}},\
  \bibinfo {pages} {425} (\bibinfo {year} {2001})},\ \Eprint
  {https://arxiv.org/abs/https://rss.onlinelibrary.wiley.com/doi/pdf/10.1111/1467-9868.00294}
  {https://rss.onlinelibrary.wiley.com/doi/pdf/10.1111/1467-9868.00294}
  \BibitemShut {NoStop}%
\bibitem [{\citenamefont {Salter}\ \emph {et~al.}(2019)\citenamefont {Salter},
  \citenamefont {Williamson}, \citenamefont {Scinocca},\ and\ \citenamefont
  {Kharin}}]{salter_uncertainty_2019}%
  \BibitemOpen
  \bibfield  {author} {\bibinfo {author} {\bibfnamefont {J.~M.}\ \bibnamefont
  {Salter}}, \bibinfo {author} {\bibfnamefont {D.~B.}\ \bibnamefont
  {Williamson}}, \bibinfo {author} {\bibfnamefont {J.}~\bibnamefont
  {Scinocca}},\ and\ \bibinfo {author} {\bibfnamefont {V.}~\bibnamefont
  {Kharin}},\ }\bibfield  {title} {\bibinfo {title} {Uncertainty quantification
  for computer models with spatial output using calibration-optimal bases},\
  }\href {https://doi.org/10.1080/01621459.2018.1514306} {\bibfield  {journal}
  {\bibinfo  {journal} {Journal of the American Statistical Association}\
  }\textbf {\bibinfo {volume} {114}},\ \bibinfo {pages} {1800} (\bibinfo {year}
  {2019})},\ \bibinfo {note} {arXiv: 1801.08184}\BibitemShut {NoStop}%
\bibitem [{\citenamefont {Stroberg}\ \emph {et~al.}(2022)\citenamefont
  {Stroberg}, \citenamefont {Henderson}, \citenamefont {Hackman}, \citenamefont
  {Ruotsalainen}, \citenamefont {Hagen},\ and\ \citenamefont
  {Holt}}]{PhysRevC.105.034333}%
  \BibitemOpen
  \bibfield  {author} {\bibinfo {author} {\bibfnamefont {S.~R.}\ \bibnamefont
  {Stroberg}}, \bibinfo {author} {\bibfnamefont {J.}~\bibnamefont {Henderson}},
  \bibinfo {author} {\bibfnamefont {G.}~\bibnamefont {Hackman}}, \bibinfo
  {author} {\bibfnamefont {P.}~\bibnamefont {Ruotsalainen}}, \bibinfo {author}
  {\bibfnamefont {G.}~\bibnamefont {Hagen}},\ and\ \bibinfo {author}
  {\bibfnamefont {J.~D.}\ \bibnamefont {Holt}},\ }\bibfield  {title} {\bibinfo
  {title} {Systematics of $e2$ strength in the $sd$ shell with the
  valence-space in-medium similarity renormalization group},\ }\href
  {https://doi.org/10.1103/PhysRevC.105.034333} {\bibfield  {journal} {\bibinfo
   {journal} {Phys. Rev. C}\ }\textbf {\bibinfo {volume} {105}},\ \bibinfo
  {pages} {034333} (\bibinfo {year} {2022})}\BibitemShut {NoStop}%
\bibitem [{\citenamefont {Sivia}\ and\ \citenamefont
  {Skilling}(2006)}]{sivia_bayes}%
  \BibitemOpen
  \bibfield  {author} {\bibinfo {author} {\bibfnamefont {D.~S.}\ \bibnamefont
  {Sivia}}\ and\ \bibinfo {author} {\bibfnamefont {J.}~\bibnamefont
  {Skilling}},\ }\href@noop {} {\emph {\bibinfo {title} {Data Analysis: A
  Bayesian Tutorial}}},\ \bibinfo {edition} {2nd}\ ed.\ (\bibinfo  {publisher}
  {Oxford Science Publications},\ \bibinfo {year} {2006})\BibitemShut {NoStop}%
\bibitem [{\citenamefont {Brooks}\ \emph {et~al.}(2011)\citenamefont {Brooks},
  \citenamefont {Gelman}, \citenamefont {Jones},\ and\ \citenamefont
  {Meng}}]{mcmc_handbook}%
  \BibitemOpen
  \bibfield  {author} {\bibinfo {author} {\bibfnamefont {S.}~\bibnamefont
  {Brooks}}, \bibinfo {author} {\bibfnamefont {A.}~\bibnamefont {Gelman}},
  \bibinfo {author} {\bibfnamefont {G.~L.}\ \bibnamefont {Jones}},\ and\
  \bibinfo {author} {\bibfnamefont {X.-L.}\ \bibnamefont {Meng}},\ }\href@noop
  {} {\emph {\bibinfo {title} {Handbook of Markov Chain Monte Carlo}}}\
  (\bibinfo  {publisher} {Chapman \& Hall/CRC},\ \bibinfo {year}
  {2011})\BibitemShut {NoStop}%
\bibitem [{\citenamefont {Duane}\ \emph {et~al.}(1987)\citenamefont {Duane},
  \citenamefont {Kennedy}, \citenamefont {Pendleton},\ and\ \citenamefont
  {Roweth}}]{hmc}%
  \BibitemOpen
  \bibfield  {author} {\bibinfo {author} {\bibfnamefont {S.}~\bibnamefont
  {Duane}}, \bibinfo {author} {\bibfnamefont {A.}~\bibnamefont {Kennedy}},
  \bibinfo {author} {\bibfnamefont {B.~J.}\ \bibnamefont {Pendleton}},\ and\
  \bibinfo {author} {\bibfnamefont {D.}~\bibnamefont {Roweth}},\ }\bibfield
  {title} {\bibinfo {title} {Hybrid monte carlo},\ }\href
  {https://doi.org/https://doi.org/10.1016/0370-2693(87)91197-X} {\bibfield
  {journal} {\bibinfo  {journal} {Physics Letters B}\ }\textbf {\bibinfo
  {volume} {195}},\ \bibinfo {pages} {216} (\bibinfo {year}
  {1987})}\BibitemShut {NoStop}%
\bibitem [{\citenamefont {Hoffman}\ \emph {et~al.}(2014)\citenamefont
  {Hoffman}, \citenamefont {Gelman} \emph {et~al.}}]{nuts}%
  \BibitemOpen
  \bibfield  {author} {\bibinfo {author} {\bibfnamefont {M.~D.}\ \bibnamefont
  {Hoffman}}, \bibinfo {author} {\bibfnamefont {A.}~\bibnamefont {Gelman}},
  \emph {et~al.},\ }\bibfield  {title} {\bibinfo {title} {The no-u-turn
  sampler: adaptively setting path lengths in hamiltonian monte carlo.},\
  }\href@noop {} {\bibfield  {journal} {\bibinfo  {journal} {J. Mach. Learn.
  Res.}\ }\textbf {\bibinfo {volume} {15}},\ \bibinfo {pages} {1593} (\bibinfo
  {year} {2014})}\BibitemShut {NoStop}%
\bibitem [{\citenamefont {{Foreman-Mackey}}\ \emph {et~al.}(2013)\citenamefont
  {{Foreman-Mackey}}, \citenamefont {{Hogg}}, \citenamefont {{Lang}},\ and\
  \citenamefont {{Goodman}}}]{emcee_paper}%
  \BibitemOpen
  \bibfield  {author} {\bibinfo {author} {\bibfnamefont {D.}~\bibnamefont
  {{Foreman-Mackey}}}, \bibinfo {author} {\bibfnamefont {D.~W.}\ \bibnamefont
  {{Hogg}}}, \bibinfo {author} {\bibfnamefont {D.}~\bibnamefont {{Lang}}},\
  and\ \bibinfo {author} {\bibfnamefont {J.}~\bibnamefont {{Goodman}}},\
  }\bibfield  {title} {\bibinfo {title} {{emcee: The MCMC Hammer}},\ }\href
  {https://doi.org/10.1086/670067} {\ \textbf {\bibinfo {volume} {125}},\
  \bibinfo {pages} {306} (\bibinfo {year} {2013})},\ \Eprint
  {https://arxiv.org/abs/1202.3665} {arXiv:1202.3665 [astro-ph.IM]}
  \BibitemShut {NoStop}%
\bibitem [{\citenamefont {Frame}\ \emph {et~al.}(2018)\citenamefont {Frame},
  \citenamefont {He}, \citenamefont {Ipsen}, \citenamefont {Lee}, \citenamefont
  {Lee},\ and\ \citenamefont {Rrapaj}}]{frame_ec}%
  \BibitemOpen
  \bibfield  {author} {\bibinfo {author} {\bibfnamefont {D.}~\bibnamefont
  {Frame}}, \bibinfo {author} {\bibfnamefont {R.}~\bibnamefont {He}}, \bibinfo
  {author} {\bibfnamefont {I.}~\bibnamefont {Ipsen}}, \bibinfo {author}
  {\bibfnamefont {D.}~\bibnamefont {Lee}}, \bibinfo {author} {\bibfnamefont
  {D.}~\bibnamefont {Lee}},\ and\ \bibinfo {author} {\bibfnamefont
  {E.}~\bibnamefont {Rrapaj}},\ }\bibfield  {title} {\bibinfo {title}
  {Eigenvector continuation with subspace learning},\ }\href@noop {} {\bibfield
   {journal} {\bibinfo  {journal} {Physical review letters}\ }\textbf {\bibinfo
  {volume} {121}},\ \bibinfo {pages} {032501} (\bibinfo {year}
  {2018})}\BibitemShut {NoStop}%
\bibitem [{\citenamefont {K{\"o}nig}\ \emph {et~al.}(2020)\citenamefont
  {K{\"o}nig}, \citenamefont {Ekstr{\"o}m}, \citenamefont {Hebeler},
  \citenamefont {Lee},\ and\ \citenamefont {Schwenk}}]{konig_ec}%
  \BibitemOpen
  \bibfield  {author} {\bibinfo {author} {\bibfnamefont {S.}~\bibnamefont
  {K{\"o}nig}}, \bibinfo {author} {\bibfnamefont {A.}~\bibnamefont
  {Ekstr{\"o}m}}, \bibinfo {author} {\bibfnamefont {K.}~\bibnamefont
  {Hebeler}}, \bibinfo {author} {\bibfnamefont {D.}~\bibnamefont {Lee}},\ and\
  \bibinfo {author} {\bibfnamefont {A.}~\bibnamefont {Schwenk}},\ }\bibfield
  {title} {\bibinfo {title} {Eigenvector continuation as an efficient and
  accurate emulator for uncertainty quantification},\ }\href@noop {} {\bibfield
   {journal} {\bibinfo  {journal} {Physics Letters B}\ }\textbf {\bibinfo
  {volume} {810}},\ \bibinfo {pages} {135814} (\bibinfo {year}
  {2020})}\BibitemShut {NoStop}%
\bibitem [{\citenamefont {Rasmussen}\ and\ \citenamefont
  {Williams}(2006)}]{rasmussen_gp}%
  \BibitemOpen
  \bibfield  {author} {\bibinfo {author} {\bibfnamefont {C.~E.}\ \bibnamefont
  {Rasmussen}}\ and\ \bibinfo {author} {\bibfnamefont {C.~K.~I.}\ \bibnamefont
  {Williams}},\ }\href@noop {} {\emph {\bibinfo {title} {Gaussian Processes for
  Machine Learning}}}\ (\bibinfo  {publisher} {MIT Press},\ \bibinfo {year}
  {2006})\BibitemShut {NoStop}%
\bibitem [{\citenamefont {Goodfellow}\ \emph {et~al.}(2016)\citenamefont
  {Goodfellow}, \citenamefont {Bengio},\ and\ \citenamefont
  {Courville}}]{goodfellow_deeplearning}%
  \BibitemOpen
  \bibfield  {author} {\bibinfo {author} {\bibfnamefont {I.}~\bibnamefont
  {Goodfellow}}, \bibinfo {author} {\bibfnamefont {Y.}~\bibnamefont {Bengio}},\
  and\ \bibinfo {author} {\bibfnamefont {A.}~\bibnamefont {Courville}},\
  }\href@noop {} {\emph {\bibinfo {title} {Deep Learning}}}\ (\bibinfo
  {publisher} {MIT Press},\ \bibinfo {year} {2016})\ \bibinfo {note}
  {\url{http://www.deeplearningbook.org}}\BibitemShut {NoStop}%
\bibitem [{\citenamefont {Neufcourt}\ \emph {et~al.}(2018)\citenamefont
  {Neufcourt}, \citenamefont {Cao}, \citenamefont {Nazarewicz},\ and\
  \citenamefont {Viens}}]{neufcourt_bayesian_extrapolation}%
  \BibitemOpen
  \bibfield  {author} {\bibinfo {author} {\bibfnamefont {L.}~\bibnamefont
  {Neufcourt}}, \bibinfo {author} {\bibfnamefont {Y.}~\bibnamefont {Cao}},
  \bibinfo {author} {\bibfnamefont {W.}~\bibnamefont {Nazarewicz}},\ and\
  \bibinfo {author} {\bibfnamefont {F.}~\bibnamefont {Viens}},\ }\bibfield
  {title} {\bibinfo {title} {{Bayesian approach to model-based extrapolation of
  nuclear observables}},\ }\href {https://doi.org/10.1103/PhysRevC.98.034318}
  {\bibfield  {journal} {\bibinfo  {journal} {Phys. Rev. C}\ }\textbf {\bibinfo
  {volume} {98}},\ \bibinfo {pages} {034318} (\bibinfo {year} {2018})},\
  \Eprint {https://arxiv.org/abs/1806.00552} {arXiv:1806.00552 [nucl-th]}
  \BibitemShut {NoStop}%
\bibitem [{\citenamefont {Brussard}\ and\ \citenamefont
  {Glaudemans}(1977)}]{BG77}%
  \BibitemOpen
  \bibfield  {author} {\bibinfo {author} {\bibfnamefont {P.}~\bibnamefont
  {Brussard}}\ and\ \bibinfo {author} {\bibfnamefont {P.}~\bibnamefont
  {Glaudemans}},\ }\href@noop {} {\emph {\bibinfo {title} {Shell-model
  applications in nuclear spectroscopy}}}\ (\bibinfo  {publisher}
  {North-Holland Publishing Company, Amsterdam},\ \bibinfo {year}
  {1977})\BibitemShut {NoStop}%
\bibitem [{\citenamefont {Brown}\ and\ \citenamefont
  {Wildenthal}(1988)}]{br88}%
  \BibitemOpen
  \bibfield  {author} {\bibinfo {author} {\bibfnamefont {B.~A.}\ \bibnamefont
  {Brown}}\ and\ \bibinfo {author} {\bibfnamefont {B.~H.}\ \bibnamefont
  {Wildenthal}},\ }\bibfield  {title} {\bibinfo {title} {Status of the nuclear
  shell model},\ }\href@noop {} {\bibfield  {journal} {\bibinfo  {journal}
  {{Annual Review of Nuclear and Particle Science}}\ }\textbf {\bibinfo
  {volume} {38}},\ \bibinfo {pages} {29} (\bibinfo {year} {1988})}\BibitemShut
  {NoStop}%
\bibitem [{\citenamefont {Caurier}\ \emph {et~al.}(2005)\citenamefont
  {Caurier}, \citenamefont {Martinez-Pinedo}, \citenamefont {Nowacki},
  \citenamefont {Poves},\ and\ \citenamefont {Zuker}}]{ca05}%
  \BibitemOpen
  \bibfield  {author} {\bibinfo {author} {\bibfnamefont {E.}~\bibnamefont
  {Caurier}}, \bibinfo {author} {\bibfnamefont {G.}~\bibnamefont
  {Martinez-Pinedo}}, \bibinfo {author} {\bibfnamefont {F.}~\bibnamefont
  {Nowacki}}, \bibinfo {author} {\bibfnamefont {A.}~\bibnamefont {Poves}},\
  and\ \bibinfo {author} {\bibfnamefont {A.~P.}\ \bibnamefont {Zuker}},\
  }\bibfield  {title} {\bibinfo {title} {The shell model as a unified view of
  nuclear structure},\ }\href@noop {} {\bibfield  {journal} {\bibinfo
  {journal} {{Reviews of Modern Physics}}\ }\textbf {\bibinfo {volume} {77}},\
  \bibinfo {pages} {427} (\bibinfo {year} {2005})}\BibitemShut {NoStop}%
\bibitem [{\citenamefont {Suhonen}(2007)}]{suhonen2007nucleons}%
  \BibitemOpen
  \bibfield  {author} {\bibinfo {author} {\bibfnamefont {J.}~\bibnamefont
  {Suhonen}},\ }\href@noop {} {\emph {\bibinfo {title} {From nucleons to
  nucleus: concepts of microscopic nuclear theory}}}\ (\bibinfo  {publisher}
  {Springer Science \& Business Media, Berlin},\ \bibinfo {year}
  {2007})\BibitemShut {NoStop}%
\bibitem [{\citenamefont {Navr{\'a}til}\ \emph {et~al.}(2000)\citenamefont
  {Navr{\'a}til}, \citenamefont {Vary},\ and\ \citenamefont
  {Barrett}}]{navratil2000large}%
  \BibitemOpen
  \bibfield  {author} {\bibinfo {author} {\bibfnamefont {P.}~\bibnamefont
  {Navr{\'a}til}}, \bibinfo {author} {\bibfnamefont {J.}~\bibnamefont {Vary}},\
  and\ \bibinfo {author} {\bibfnamefont {B.}~\bibnamefont {Barrett}},\
  }\bibfield  {title} {\bibinfo {title} {Large-basis ab initio no-core shell
  model and its application to 12 c},\ }\href@noop {} {\bibfield  {journal}
  {\bibinfo  {journal} {Physical Review C}\ }\textbf {\bibinfo {volume} {62}},\
  \bibinfo {pages} {054311} (\bibinfo {year} {2000})}\BibitemShut {NoStop}%
\bibitem [{\citenamefont {Barrett}\ \emph {et~al.}(2013)\citenamefont
  {Barrett}, \citenamefont {Navr{\'a}til},\ and\ \citenamefont
  {Vary}}]{barrett2013ab}%
  \BibitemOpen
  \bibfield  {author} {\bibinfo {author} {\bibfnamefont {B.~R.}\ \bibnamefont
  {Barrett}}, \bibinfo {author} {\bibfnamefont {P.}~\bibnamefont
  {Navr{\'a}til}},\ and\ \bibinfo {author} {\bibfnamefont {J.~P.}\ \bibnamefont
  {Vary}},\ }\bibfield  {title} {\bibinfo {title} {Ab initio no core shell
  model},\ }\href@noop {} {\bibfield  {journal} {\bibinfo  {journal} {Progress
  in Particle and Nuclear Physics}\ }\textbf {\bibinfo {volume} {69}},\
  \bibinfo {pages} {131} (\bibinfo {year} {2013})}\BibitemShut {NoStop}%
\bibitem [{\citenamefont {van Kolck}(1994)}]{PhysRevC.49.2932}%
  \BibitemOpen
  \bibfield  {author} {\bibinfo {author} {\bibfnamefont {U.}~\bibnamefont {van
  Kolck}},\ }\bibfield  {title} {\bibinfo {title} {Few-nucleon forces from
  chiral lagrangians},\ }\href@noop {} {\bibfield  {journal} {\bibinfo
  {journal} {Phys. Rev. C}\ }\textbf {\bibinfo {volume} {49}},\ \bibinfo
  {pages} {2932} (\bibinfo {year} {1994})}\BibitemShut {NoStop}%
\bibitem [{\citenamefont {Wendt}\ \emph {et~al.}(2014)\citenamefont {Wendt},
  \citenamefont {Carlsson},\ and\ \citenamefont
  {Ekstr{\"o}m}}]{wendt2014uncertainty}%
  \BibitemOpen
  \bibfield  {author} {\bibinfo {author} {\bibfnamefont {K.}~\bibnamefont
  {Wendt}}, \bibinfo {author} {\bibfnamefont {B.}~\bibnamefont {Carlsson}},\
  and\ \bibinfo {author} {\bibfnamefont {A.}~\bibnamefont {Ekstr{\"o}m}},\
  }\bibfield  {title} {\bibinfo {title} {Uncertainty quantification of the
  pion-nucleon low-energy coupling constants up to fourth order in chiral
  perturbation theory},\ }\href@noop {} {\bibfield  {journal} {\bibinfo
  {journal} {arXiv preprint arXiv:1410.0646}\ } (\bibinfo {year}
  {2014})}\BibitemShut {NoStop}%
\bibitem [{\citenamefont {Furnstahl}\ \emph
  {et~al.}(2015{\natexlab{a}})\citenamefont {Furnstahl}, \citenamefont
  {Phillips},\ and\ \citenamefont {Wesolowski}}]{furnstahl2015recipe}%
  \BibitemOpen
  \bibfield  {author} {\bibinfo {author} {\bibfnamefont {R.}~\bibnamefont
  {Furnstahl}}, \bibinfo {author} {\bibfnamefont {D.}~\bibnamefont
  {Phillips}},\ and\ \bibinfo {author} {\bibfnamefont {S.}~\bibnamefont
  {Wesolowski}},\ }\bibfield  {title} {\bibinfo {title} {A recipe for eft
  uncertainty quantification in nuclear physics},\ }\href@noop {} {\bibfield
  {journal} {\bibinfo  {journal} {Journal of Physics G: Nuclear and Particle
  Physics}\ }\textbf {\bibinfo {volume} {42}},\ \bibinfo {pages} {034028}
  (\bibinfo {year} {2015}{\natexlab{a}})}\BibitemShut {NoStop}%
\bibitem [{\citenamefont {Carlsson}\ \emph {et~al.}(2016)\citenamefont
  {Carlsson}, \citenamefont {Ekstr{\"o}m}, \citenamefont {Forss{\'e}n},
  \citenamefont {Str{\"o}mberg}, \citenamefont {Jansen}, \citenamefont {Lilja},
  \citenamefont {Lindby}, \citenamefont {Mattsson},\ and\ \citenamefont
  {Wendt}}]{carlsson2016uncertainty}%
  \BibitemOpen
  \bibfield  {author} {\bibinfo {author} {\bibfnamefont {B.}~\bibnamefont
  {Carlsson}}, \bibinfo {author} {\bibfnamefont {A.}~\bibnamefont
  {Ekstr{\"o}m}}, \bibinfo {author} {\bibfnamefont {C.}~\bibnamefont
  {Forss{\'e}n}}, \bibinfo {author} {\bibfnamefont {D.~F.}\ \bibnamefont
  {Str{\"o}mberg}}, \bibinfo {author} {\bibfnamefont {G.}~\bibnamefont
  {Jansen}}, \bibinfo {author} {\bibfnamefont {O.}~\bibnamefont {Lilja}},
  \bibinfo {author} {\bibfnamefont {M.}~\bibnamefont {Lindby}}, \bibinfo
  {author} {\bibfnamefont {B.}~\bibnamefont {Mattsson}},\ and\ \bibinfo
  {author} {\bibfnamefont {K.}~\bibnamefont {Wendt}},\ }\bibfield  {title}
  {\bibinfo {title} {Uncertainty analysis and order-by-order optimization of
  chiral nuclear interactions},\ }\href@noop {} {\bibfield  {journal} {\bibinfo
   {journal} {Physical Review X}\ }\textbf {\bibinfo {volume} {6}},\ \bibinfo
  {pages} {011019} (\bibinfo {year} {2016})}\BibitemShut {NoStop}%
\bibitem [{\citenamefont {P{\'e}rez}\ \emph {et~al.}(2016)\citenamefont
  {P{\'e}rez}, \citenamefont {Amaro},\ and\ \citenamefont
  {Arriola}}]{perez2016uncertainty}%
  \BibitemOpen
  \bibfield  {author} {\bibinfo {author} {\bibfnamefont {R.~N.}\ \bibnamefont
  {P{\'e}rez}}, \bibinfo {author} {\bibfnamefont {J.}~\bibnamefont {Amaro}},\
  and\ \bibinfo {author} {\bibfnamefont {E.~R.}\ \bibnamefont {Arriola}},\
  }\bibfield  {title} {\bibinfo {title} {Uncertainty quantification of
  effective nuclear interactions},\ }\href@noop {} {\bibfield  {journal}
  {\bibinfo  {journal} {International Journal of Modern Physics E}\ }\textbf
  {\bibinfo {volume} {25}},\ \bibinfo {pages} {1641009} (\bibinfo {year}
  {2016})}\BibitemShut {NoStop}%
\bibitem [{\citenamefont {Melendez}\ \emph {et~al.}(2019)\citenamefont
  {Melendez}, \citenamefont {Furnstahl}, \citenamefont {Phillips},
  \citenamefont {Pratola},\ and\ \citenamefont
  {Wesolowski}}]{PhysRevC.100.044001}%
  \BibitemOpen
  \bibfield  {author} {\bibinfo {author} {\bibfnamefont {J.~A.}\ \bibnamefont
  {Melendez}}, \bibinfo {author} {\bibfnamefont {R.~J.}\ \bibnamefont
  {Furnstahl}}, \bibinfo {author} {\bibfnamefont {D.~R.}\ \bibnamefont
  {Phillips}}, \bibinfo {author} {\bibfnamefont {M.~T.}\ \bibnamefont
  {Pratola}},\ and\ \bibinfo {author} {\bibfnamefont {S.}~\bibnamefont
  {Wesolowski}},\ }\bibfield  {title} {\bibinfo {title} {Quantifying correlated
  truncation errors in effective field theory},\ }\href@noop {} {\bibfield
  {journal} {\bibinfo  {journal} {Phys. Rev. C}\ }\textbf {\bibinfo {volume}
  {100}},\ \bibinfo {pages} {044001} (\bibinfo {year} {2019})}\BibitemShut
  {NoStop}%
\bibitem [{\citenamefont {Wesolowski}\ \emph {et~al.}(2019)\citenamefont
  {Wesolowski}, \citenamefont {Furnstahl}, \citenamefont {Melendez},\ and\
  \citenamefont {Phillips}}]{wesolowski2019exploring}%
  \BibitemOpen
  \bibfield  {author} {\bibinfo {author} {\bibfnamefont {S.}~\bibnamefont
  {Wesolowski}}, \bibinfo {author} {\bibfnamefont {R.}~\bibnamefont
  {Furnstahl}}, \bibinfo {author} {\bibfnamefont {J.}~\bibnamefont
  {Melendez}},\ and\ \bibinfo {author} {\bibfnamefont {D.}~\bibnamefont
  {Phillips}},\ }\bibfield  {title} {\bibinfo {title} {Exploring bayesian
  parameter estimation for chiral effective field theory using nucleon--nucleon
  phase shifts},\ }\href@noop {} {\bibfield  {journal} {\bibinfo  {journal}
  {Journal of Physics G: Nuclear and Particle Physics}\ }\textbf {\bibinfo
  {volume} {46}},\ \bibinfo {pages} {045102} (\bibinfo {year}
  {2019})}\BibitemShut {NoStop}%
\bibitem [{\citenamefont {Stroberg}\ \emph {et~al.}(2019)\citenamefont
  {Stroberg}, \citenamefont {Hergert}, \citenamefont {Bogner},\ and\
  \citenamefont {Holt}}]{stroberg2019nonempirical}%
  \BibitemOpen
  \bibfield  {author} {\bibinfo {author} {\bibfnamefont {S.~R.}\ \bibnamefont
  {Stroberg}}, \bibinfo {author} {\bibfnamefont {H.}~\bibnamefont {Hergert}},
  \bibinfo {author} {\bibfnamefont {S.~K.}\ \bibnamefont {Bogner}},\ and\
  \bibinfo {author} {\bibfnamefont {J.~D.}\ \bibnamefont {Holt}},\ }\bibfield
  {title} {\bibinfo {title} {Nonempirical interactions for the nuclear shell
  model: an update},\ }\href@noop {} {\bibfield  {journal} {\bibinfo  {journal}
  {Annual Review of Nuclear and Particle Science}\ }\textbf {\bibinfo {volume}
  {69}},\ \bibinfo {pages} {307} (\bibinfo {year} {2019})}\BibitemShut
  {NoStop}%
\bibitem [{\citenamefont {Bogner}\ \emph {et~al.}(2007)\citenamefont {Bogner},
  \citenamefont {Furnstahl},\ and\ \citenamefont {Perry}}]{PhysRevC.75.061001}%
  \BibitemOpen
  \bibfield  {author} {\bibinfo {author} {\bibfnamefont {S.~K.}\ \bibnamefont
  {Bogner}}, \bibinfo {author} {\bibfnamefont {R.~J.}\ \bibnamefont
  {Furnstahl}},\ and\ \bibinfo {author} {\bibfnamefont {R.~J.}\ \bibnamefont
  {Perry}},\ }\bibfield  {title} {\bibinfo {title} {Similarity renormalization
  group for nucleon-nucleon interactions},\ }\href
  {https://doi.org/10.1103/PhysRevC.75.061001} {\bibfield  {journal} {\bibinfo
  {journal} {Phys. Rev. C}\ }\textbf {\bibinfo {volume} {75}},\ \bibinfo
  {pages} {061001} (\bibinfo {year} {2007})}\BibitemShut {NoStop}%
\bibitem [{\citenamefont {Hergert}\ \emph {et~al.}(2016)\citenamefont
  {Hergert}, \citenamefont {Bogner}, \citenamefont {Morris}, \citenamefont
  {Schwenk},\ and\ \citenamefont {Tsukiyama}}]{hergert2016medium}%
  \BibitemOpen
  \bibfield  {author} {\bibinfo {author} {\bibfnamefont {H.}~\bibnamefont
  {Hergert}}, \bibinfo {author} {\bibfnamefont {S.}~\bibnamefont {Bogner}},
  \bibinfo {author} {\bibfnamefont {T.}~\bibnamefont {Morris}}, \bibinfo
  {author} {\bibfnamefont {A.}~\bibnamefont {Schwenk}},\ and\ \bibinfo {author}
  {\bibfnamefont {K.}~\bibnamefont {Tsukiyama}},\ }\bibfield  {title} {\bibinfo
  {title} {The in-medium similarity renormalization group: A novel ab initio
  method for nuclei},\ }\href@noop {} {\bibfield  {journal} {\bibinfo
  {journal} {Physics Reports}\ }\textbf {\bibinfo {volume} {621}},\ \bibinfo
  {pages} {165} (\bibinfo {year} {2016})}\BibitemShut {NoStop}%
\bibitem [{\citenamefont {Johnson}\ \emph {et~al.}(2013)\citenamefont
  {Johnson}, \citenamefont {Ormand},\ and\ \citenamefont {Krastev}}]{BIGSTICK}%
  \BibitemOpen
  \bibfield  {author} {\bibinfo {author} {\bibfnamefont {C.~W.}\ \bibnamefont
  {Johnson}}, \bibinfo {author} {\bibfnamefont {W.~E.}\ \bibnamefont
  {Ormand}},\ and\ \bibinfo {author} {\bibfnamefont {P.~G.}\ \bibnamefont
  {Krastev}},\ }\bibfield  {title} {\bibinfo {title} {Factorization in
  large-scale many-body calculations},\ }\href@noop {} {\bibfield  {journal}
  {\bibinfo  {journal} {Computer Physics Communications}\ }\textbf {\bibinfo
  {volume} {184}},\ \bibinfo {pages} {2761} (\bibinfo {year}
  {2013})}\BibitemShut {NoStop}%
\bibitem [{\citenamefont {Johnson}\ \emph {et~al.}(2018)\citenamefont
  {Johnson}, \citenamefont {Ormand}, \citenamefont {McElvain},\ and\
  \citenamefont {Shan}}]{johnson2018bigstick}%
  \BibitemOpen
  \bibfield  {author} {\bibinfo {author} {\bibfnamefont {C.~W.}\ \bibnamefont
  {Johnson}}, \bibinfo {author} {\bibfnamefont {W.~E.}\ \bibnamefont {Ormand}},
  \bibinfo {author} {\bibfnamefont {K.~S.}\ \bibnamefont {McElvain}},\ and\
  \bibinfo {author} {\bibfnamefont {H.}~\bibnamefont {Shan}},\ }\bibfield
  {title} {\bibinfo {title} {Bigstick: A flexible configuration-interaction
  shell-model code},\ }\href@noop {} {\bibfield  {journal} {\bibinfo  {journal}
  {arXiv preprint arXiv:1801.08432}\ } (\bibinfo {year} {2018})}\BibitemShut
  {NoStop}%
\bibitem [{\citenamefont {Brown}\ and\ \citenamefont
  {Richter}(2006)}]{brown_usd_hams}%
  \BibitemOpen
  \bibfield  {author} {\bibinfo {author} {\bibfnamefont {B.~A.}\ \bibnamefont
  {Brown}}\ and\ \bibinfo {author} {\bibfnamefont {W.~A.}\ \bibnamefont
  {Richter}},\ }\bibfield  {title} {\bibinfo {title} {New ``usd'' hamiltonians
  for the $\mathit{sd}$ shell},\ }\href
  {https://doi.org/10.1103/PhysRevC.74.034315} {\bibfield  {journal} {\bibinfo
  {journal} {Phys. Rev. C}\ }\textbf {\bibinfo {volume} {74}},\ \bibinfo
  {pages} {034315} (\bibinfo {year} {2006})}\BibitemShut {NoStop}%
\bibitem [{\citenamefont {Fox}\ \emph {et~al.}(2020)\citenamefont {Fox},
  \citenamefont {Johnson},\ and\ \citenamefont {Perez}}]{fox_usdb}%
  \BibitemOpen
  \bibfield  {author} {\bibinfo {author} {\bibfnamefont {J.~M.~R.}\
  \bibnamefont {Fox}}, \bibinfo {author} {\bibfnamefont {C.~W.}\ \bibnamefont
  {Johnson}},\ and\ \bibinfo {author} {\bibfnamefont {R.~N.}\ \bibnamefont
  {Perez}},\ }\bibfield  {title} {\bibinfo {title} {Uncertainty quantification
  of an empirical shell-model interaction using principal component analysis},\
  }\bibfield  {journal} {\bibinfo  {journal} {Physical Review C}\ }\textbf
  {\bibinfo {volume} {101}},\ \href
  {https://doi.org/10.1103/physrevc.101.054308} {10.1103/physrevc.101.054308}
  (\bibinfo {year} {2020})\BibitemShut {NoStop}%
\bibitem [{\citenamefont {Ghanem}\ \emph {et~al.}(2017)\citenamefont {Ghanem},
  \citenamefont {Higdon},\ and\ \citenamefont {Owhadi}}]{ghanem2017handbook}%
  \BibitemOpen
  \bibfield  {author} {\bibinfo {author} {\bibfnamefont {R.}~\bibnamefont
  {Ghanem}}, \bibinfo {author} {\bibfnamefont {D.}~\bibnamefont {Higdon}},\
  and\ \bibinfo {author} {\bibfnamefont {e.~a.}\ \bibnamefont {Owhadi},
  \bibfnamefont {Houman}},\ }\href@noop {} {\emph {\bibinfo {title} {Handbook
  of uncertainty quantification}}},\ Vol.~\bibinfo {volume} {6}\ (\bibinfo
  {publisher} {Springer, New York},\ \bibinfo {year} {2017})\BibitemShut
  {NoStop}%
\bibitem [{\citenamefont {Hellman}(1937)}]{hellman1937einfuhrung}%
  \BibitemOpen
  \bibfield  {author} {\bibinfo {author} {\bibfnamefont {H.}~\bibnamefont
  {Hellman}},\ }\bibfield  {title} {\bibinfo {title} {Einf{\"u}hrung in die
  quantenchemie},\ }\href@noop {} {\bibfield  {journal} {\bibinfo  {journal}
  {Franz Deuticke, Leipzig}\ ,\ \bibinfo {pages} {285}} (\bibinfo {year}
  {1937})}\BibitemShut {NoStop}%
\bibitem [{\citenamefont {Feynman}(1939)}]{PhysRev.56.340}%
  \BibitemOpen
  \bibfield  {author} {\bibinfo {author} {\bibfnamefont {R.~P.}\ \bibnamefont
  {Feynman}},\ }\bibfield  {title} {\bibinfo {title} {Forces in molecules},\
  }\href {https://doi.org/10.1103/PhysRev.56.340} {\bibfield  {journal}
  {\bibinfo  {journal} {Phys. Rev.}\ }\textbf {\bibinfo {volume} {56}},\
  \bibinfo {pages} {340} (\bibinfo {year} {1939})}\BibitemShut {NoStop}%
\bibitem [{\citenamefont {Yoshida}\ and\ \citenamefont
  {Shimizu}(2021)}]{yoshida_eigenvector_continuation}%
  \BibitemOpen
  \bibfield  {author} {\bibinfo {author} {\bibfnamefont {S.}~\bibnamefont
  {Yoshida}}\ and\ \bibinfo {author} {\bibfnamefont {N.}~\bibnamefont
  {Shimizu}},\ }\bibfield  {title} {\bibinfo {title} {{Constructing approximate
  shell-model wavefunctions by eigenvector continuation}}\ }\href
  {https://doi.org/10.1093/ptep/ptac057} {10.1093/ptep/ptac057} (\bibinfo
  {year} {2021}),\ \Eprint {https://arxiv.org/abs/2105.08256} {arXiv:2105.08256
  [nucl-th]} \BibitemShut {NoStop}%
\bibitem [{\citenamefont {Edmonds}(1996)}]{edmonds1996angular}%
  \BibitemOpen
  \bibfield  {author} {\bibinfo {author} {\bibfnamefont {A.~R.}\ \bibnamefont
  {Edmonds}},\ }\href@noop {} {\emph {\bibinfo {title} {Angular momentum in
  quantum mechanics}}}\ (\bibinfo  {publisher} {Princeton University Press,
  Princeton},\ \bibinfo {year} {1996})\BibitemShut {NoStop}%
\bibitem [{\citenamefont {Richter}\ \emph {et~al.}(2008)\citenamefont
  {Richter}, \citenamefont {Mkhize},\ and\ \citenamefont
  {Brown}}]{richter_sd_obs}%
  \BibitemOpen
  \bibfield  {author} {\bibinfo {author} {\bibfnamefont {W.~A.}\ \bibnamefont
  {Richter}}, \bibinfo {author} {\bibfnamefont {S.}~\bibnamefont {Mkhize}},\
  and\ \bibinfo {author} {\bibfnamefont {B.~A.}\ \bibnamefont {Brown}},\
  }\bibfield  {title} {\bibinfo {title} {$\mathit{sd}$-shell observables for
  the usda and usdb hamiltonians},\ }\href
  {https://doi.org/10.1103/PhysRevC.78.064302} {\bibfield  {journal} {\bibinfo
  {journal} {Phys. Rev. C}\ }\textbf {\bibinfo {volume} {78}},\ \bibinfo
  {pages} {064302} (\bibinfo {year} {2008})}\BibitemShut {NoStop}%
\bibitem [{\citenamefont {Gysbers}\ \emph {et~al.}(2019)\citenamefont
  {Gysbers}, \citenamefont {Hagen}, \citenamefont {Holt}, \citenamefont
  {Jansen}, \citenamefont {Morris}, \citenamefont {Navr{\'a}til}, \citenamefont
  {Papenbrock}, \citenamefont {Quaglioni}, \citenamefont {Schwenk},
  \citenamefont {Stroberg} \emph {et~al.}}]{gysbers2019discrepancy}%
  \BibitemOpen
  \bibfield  {author} {\bibinfo {author} {\bibfnamefont {P.}~\bibnamefont
  {Gysbers}}, \bibinfo {author} {\bibfnamefont {G.}~\bibnamefont {Hagen}},
  \bibinfo {author} {\bibfnamefont {J.}~\bibnamefont {Holt}}, \bibinfo {author}
  {\bibfnamefont {G.~R.}\ \bibnamefont {Jansen}}, \bibinfo {author}
  {\bibfnamefont {T.~D.}\ \bibnamefont {Morris}}, \bibinfo {author}
  {\bibfnamefont {P.}~\bibnamefont {Navr{\'a}til}}, \bibinfo {author}
  {\bibfnamefont {T.}~\bibnamefont {Papenbrock}}, \bibinfo {author}
  {\bibfnamefont {S.}~\bibnamefont {Quaglioni}}, \bibinfo {author}
  {\bibfnamefont {A.}~\bibnamefont {Schwenk}}, \bibinfo {author} {\bibfnamefont
  {S.}~\bibnamefont {Stroberg}}, \emph {et~al.},\ }\bibfield  {title} {\bibinfo
  {title} {Discrepancy between experimental and theoretical $\beta$-decay rates
  resolved from first principles},\ }\href@noop {} {\bibfield  {journal}
  {\bibinfo  {journal} {Nature Physics}\ }\textbf {\bibinfo {volume} {15}},\
  \bibinfo {pages} {428} (\bibinfo {year} {2019})}\BibitemShut {NoStop}%
\bibitem [{\citenamefont {Dronchi}\ \emph {et~al.}(2023)\citenamefont
  {Dronchi}, \citenamefont {Weisshaar}, \citenamefont {Brown}, \citenamefont
  {Gade}, \citenamefont {Charity}, \citenamefont {Sobotka}, \citenamefont
  {Brown}, \citenamefont {Reviol}, \citenamefont {Bazin}, \citenamefont
  {Farris}, \citenamefont {Hill}, \citenamefont {Li}, \citenamefont
  {Longfellow}, \citenamefont {Rhodes}, \citenamefont {Paneru}, \citenamefont
  {Gillespie}, \citenamefont {Anthony}, \citenamefont {Rubino},\ and\
  \citenamefont {Biswas}}]{PhysRevC.107.034306}%
  \BibitemOpen
  \bibfield  {author} {\bibinfo {author} {\bibfnamefont {N.}~\bibnamefont
  {Dronchi}}, \bibinfo {author} {\bibfnamefont {D.}~\bibnamefont {Weisshaar}},
  \bibinfo {author} {\bibfnamefont {B.~A.}\ \bibnamefont {Brown}}, \bibinfo
  {author} {\bibfnamefont {A.}~\bibnamefont {Gade}}, \bibinfo {author}
  {\bibfnamefont {R.~J.}\ \bibnamefont {Charity}}, \bibinfo {author}
  {\bibfnamefont {L.~G.}\ \bibnamefont {Sobotka}}, \bibinfo {author}
  {\bibfnamefont {K.~W.}\ \bibnamefont {Brown}}, \bibinfo {author}
  {\bibfnamefont {W.}~\bibnamefont {Reviol}}, \bibinfo {author} {\bibfnamefont
  {D.}~\bibnamefont {Bazin}}, \bibinfo {author} {\bibfnamefont {P.~J.}\
  \bibnamefont {Farris}}, \bibinfo {author} {\bibfnamefont {A.~M.}\
  \bibnamefont {Hill}}, \bibinfo {author} {\bibfnamefont {J.}~\bibnamefont
  {Li}}, \bibinfo {author} {\bibfnamefont {B.}~\bibnamefont {Longfellow}},
  \bibinfo {author} {\bibfnamefont {D.}~\bibnamefont {Rhodes}}, \bibinfo
  {author} {\bibfnamefont {S.~N.}\ \bibnamefont {Paneru}}, \bibinfo {author}
  {\bibfnamefont {S.~A.}\ \bibnamefont {Gillespie}}, \bibinfo {author}
  {\bibfnamefont {A.}~\bibnamefont {Anthony}}, \bibinfo {author} {\bibfnamefont
  {E.}~\bibnamefont {Rubino}},\ and\ \bibinfo {author} {\bibfnamefont
  {S.}~\bibnamefont {Biswas}},\ }\bibfield  {title} {\bibinfo {title}
  {Measurement of the $b(e2\ensuremath{\uparrow})$ strengths of
  $^{36}\mathrm{Ca}$ and $^{38}\mathrm{Ca}$},\ }\href
  {https://doi.org/10.1103/PhysRevC.107.034306} {\bibfield  {journal} {\bibinfo
   {journal} {Phys. Rev. C}\ }\textbf {\bibinfo {volume} {107}},\ \bibinfo
  {pages} {034306} (\bibinfo {year} {2023})}\BibitemShut {NoStop}%
\bibitem [{\citenamefont {Navarro~P{\'e}rez}\ \emph {et~al.}(2014)\citenamefont
  {Navarro~P{\'e}rez}, \citenamefont {Amaro},\ and\ \citenamefont
  {Ruiz~Arriola}}]{navarroperez_Bootstrapping_2014}%
  \BibitemOpen
  \bibfield  {author} {\bibinfo {author} {\bibfnamefont {R.}~\bibnamefont
  {Navarro~P{\'e}rez}}, \bibinfo {author} {\bibfnamefont {J.~E.}\ \bibnamefont
  {Amaro}},\ and\ \bibinfo {author} {\bibfnamefont {E.}~\bibnamefont
  {Ruiz~Arriola}},\ }\bibfield  {title} {\bibinfo {title} {Bootstrapping the
  statistical uncertainties of {{NN}} scattering data},\ }\href
  {https://doi.org/10.1016/j.physletb.2014.09.035} {\bibfield  {journal}
  {\bibinfo  {journal} {Physics Letters B}\ }\textbf {\bibinfo {volume}
  {738}},\ \bibinfo {pages} {155} (\bibinfo {year} {2014})}\BibitemShut
  {NoStop}%
\bibitem [{\citenamefont {Binder}\ \emph {et~al.}(2014)\citenamefont {Binder},
  \citenamefont {Langhammer}, \citenamefont {Calci},\ and\ \citenamefont
  {Roth}}]{binder_Initio_2014}%
  \BibitemOpen
  \bibfield  {author} {\bibinfo {author} {\bibfnamefont {S.}~\bibnamefont
  {Binder}}, \bibinfo {author} {\bibfnamefont {J.}~\bibnamefont {Langhammer}},
  \bibinfo {author} {\bibfnamefont {A.}~\bibnamefont {Calci}},\ and\ \bibinfo
  {author} {\bibfnamefont {R.}~\bibnamefont {Roth}},\ }\bibfield  {title}
  {\bibinfo {title} {{\emph{Ab Initio}} path to heavy nuclei},\ }\href
  {https://doi.org/10.1016/j.physletb.2014.07.010} {\bibfield  {journal}
  {\bibinfo  {journal} {Physics Letters B}\ }\textbf {\bibinfo {volume}
  {736}},\ \bibinfo {pages} {119} (\bibinfo {year} {2014})}\BibitemShut
  {NoStop}%
\bibitem [{\citenamefont {Furnstahl}\ \emph
  {et~al.}(2015{\natexlab{b}})\citenamefont {Furnstahl}, \citenamefont {Hagen},
  \citenamefont {Papenbrock},\ and\ \citenamefont
  {Wendt}}]{furnstahl_Infrared_2015}%
  \BibitemOpen
  \bibfield  {author} {\bibinfo {author} {\bibfnamefont {R.~J.}\ \bibnamefont
  {Furnstahl}}, \bibinfo {author} {\bibfnamefont {G.}~\bibnamefont {Hagen}},
  \bibinfo {author} {\bibfnamefont {T.}~\bibnamefont {Papenbrock}},\ and\
  \bibinfo {author} {\bibfnamefont {K.~A.}\ \bibnamefont {Wendt}},\ }\bibfield
  {title} {\bibinfo {title} {Infrared extrapolations for atomic nuclei},\
  }\href {https://doi.org/10.1088/0954-3899/42/3/034032} {\bibfield  {journal}
  {\bibinfo  {journal} {Journal of Physics G: Nuclear and Particle Physics}\
  }\textbf {\bibinfo {volume} {42}},\ \bibinfo {pages} {034032} (\bibinfo
  {year} {2015}{\natexlab{b}})}\BibitemShut {NoStop}%
\bibitem [{\citenamefont {Wendt}\ \emph {et~al.}(2015)\citenamefont {Wendt},
  \citenamefont {Forss{\'e}n}, \citenamefont {Papenbrock},\ and\ \citenamefont
  {S{\"a}{\"a}f}}]{wendt_Infrared_2015}%
  \BibitemOpen
  \bibfield  {author} {\bibinfo {author} {\bibfnamefont {K.~A.}\ \bibnamefont
  {Wendt}}, \bibinfo {author} {\bibfnamefont {C.}~\bibnamefont {Forss{\'e}n}},
  \bibinfo {author} {\bibfnamefont {T.}~\bibnamefont {Papenbrock}},\ and\
  \bibinfo {author} {\bibfnamefont {D.}~\bibnamefont {S{\"a}{\"a}f}},\
  }\bibfield  {title} {\bibinfo {title} {Infrared length scale and
  extrapolations for the no-core shell model},\ }\href
  {https://doi.org/10.1103/PhysRevC.91.061301} {\bibfield  {journal} {\bibinfo
  {journal} {Physical Review C}\ }\textbf {\bibinfo {volume} {91}},\ \bibinfo
  {pages} {061301} (\bibinfo {year} {2015})}\BibitemShut {NoStop}%
\bibitem [{\citenamefont {Ekstr{\"o}m}\ \emph {et~al.}(2019)\citenamefont
  {Ekstr{\"o}m}, \citenamefont {Forss{\'e}n}, \citenamefont {Dimitrakakis},
  \citenamefont {Dubhashi}, \citenamefont {Johansson}, \citenamefont
  {Muhammad}, \citenamefont {Salomonsson},\ and\ \citenamefont
  {Schliep}}]{ekstrom_Bayesian_2019}%
  \BibitemOpen
  \bibfield  {author} {\bibinfo {author} {\bibfnamefont {A.}~\bibnamefont
  {Ekstr{\"o}m}}, \bibinfo {author} {\bibfnamefont {C.}~\bibnamefont
  {Forss{\'e}n}}, \bibinfo {author} {\bibfnamefont {C.}~\bibnamefont
  {Dimitrakakis}}, \bibinfo {author} {\bibfnamefont {D.}~\bibnamefont
  {Dubhashi}}, \bibinfo {author} {\bibfnamefont {H.~T.}\ \bibnamefont
  {Johansson}}, \bibinfo {author} {\bibfnamefont {A.~S.}\ \bibnamefont
  {Muhammad}}, \bibinfo {author} {\bibfnamefont {H.}~\bibnamefont
  {Salomonsson}},\ and\ \bibinfo {author} {\bibfnamefont {A.}~\bibnamefont
  {Schliep}},\ }\bibfield  {title} {\bibinfo {title} {Bayesian optimization in
  {\emph{ab initio}} nuclear physics},\ }\href@noop {} {\bibfield  {journal}
  {\bibinfo  {journal} {arXiv:1902.00941 [nucl-th, stat]}\ } (\bibinfo {year}
  {2019})},\ \Eprint {https://arxiv.org/abs/1902.00941} {1902.00941 [nucl-th,
  stat]} \BibitemShut {NoStop}%
\bibitem [{\citenamefont {{Roca-Maza}}\ \emph {et~al.}(2015)\citenamefont
  {{Roca-Maza}}, \citenamefont {Paar},\ and\ \citenamefont
  {Col{\`o}}}]{roca-maza_Covariance_2015}%
  \BibitemOpen
  \bibfield  {author} {\bibinfo {author} {\bibfnamefont {X.}~\bibnamefont
  {{Roca-Maza}}}, \bibinfo {author} {\bibfnamefont {N.}~\bibnamefont {Paar}},\
  and\ \bibinfo {author} {\bibfnamefont {G.}~\bibnamefont {Col{\`o}}},\
  }\bibfield  {title} {\bibinfo {title} {Covariance analysis for energy density
  functionals and instabilities},\ }\href
  {https://doi.org/10.1088/0954-3899/42/3/034033} {\bibfield  {journal}
  {\bibinfo  {journal} {Journal of Physics G: Nuclear and Particle Physics}\
  }\textbf {\bibinfo {volume} {42}},\ \bibinfo {pages} {034033} (\bibinfo
  {year} {2015})}\BibitemShut {NoStop}%
\bibitem [{\citenamefont {Schunck}\ \emph {et~al.}(2015)\citenamefont
  {Schunck}, \citenamefont {McDonnell}, \citenamefont {Sarich}, \citenamefont
  {Wild},\ and\ \citenamefont {Higdon}}]{schunckErrorAnalysisNuclear2015}%
  \BibitemOpen
  \bibfield  {author} {\bibinfo {author} {\bibfnamefont {N.}~\bibnamefont
  {Schunck}}, \bibinfo {author} {\bibfnamefont {J.~D.}\ \bibnamefont
  {McDonnell}}, \bibinfo {author} {\bibfnamefont {J.}~\bibnamefont {Sarich}},
  \bibinfo {author} {\bibfnamefont {S.~M.}\ \bibnamefont {Wild}},\ and\
  \bibinfo {author} {\bibfnamefont {D.}~\bibnamefont {Higdon}},\ }\bibfield
  {title} {\bibinfo {title} {Error analysis in nuclear density functional
  theory},\ }\href {https://doi.org/10.1088/0954-3899/42/3/034024} {\bibfield
  {journal} {\bibinfo  {journal} {Journal of Physics G: Nuclear and Particle
  Physics}\ }\textbf {\bibinfo {volume} {42}},\ \bibinfo {pages} {034024}
  (\bibinfo {year} {2015})}\BibitemShut {NoStop}%
\bibitem [{\citenamefont {Erler}\ and\ \citenamefont
  {Reinhard}(2015)}]{erler_Error_2015}%
  \BibitemOpen
  \bibfield  {author} {\bibinfo {author} {\bibfnamefont {J.}~\bibnamefont
  {Erler}}\ and\ \bibinfo {author} {\bibfnamefont {P.-G.}\ \bibnamefont
  {Reinhard}},\ }\bibfield  {title} {\bibinfo {title} {Error estimates for the
  {{Skyrme}}\textendash{{Hartree}}\textendash{{Fock}} model},\ }\href
  {https://doi.org/10.1088/0954-3899/42/3/034026} {\bibfield  {journal}
  {\bibinfo  {journal} {Journal of Physics G: Nuclear and Particle Physics}\
  }\textbf {\bibinfo {volume} {42}},\ \bibinfo {pages} {034026} (\bibinfo
  {year} {2015})}\BibitemShut {NoStop}%
\bibitem [{\citenamefont {Neufcourt}\ \emph {et~al.}(2020)\citenamefont
  {Neufcourt}, \citenamefont {Cao}, \citenamefont {Giuliani}, \citenamefont
  {Nazarewicz}, \citenamefont {Olsen},\ and\ \citenamefont
  {Tarasov}}]{neufcourtQuantifiedLimitsNuclear2020}%
  \BibitemOpen
  \bibfield  {author} {\bibinfo {author} {\bibfnamefont {L.}~\bibnamefont
  {Neufcourt}}, \bibinfo {author} {\bibfnamefont {Y.}~\bibnamefont {Cao}},
  \bibinfo {author} {\bibfnamefont {S.~A.}\ \bibnamefont {Giuliani}}, \bibinfo
  {author} {\bibfnamefont {W.}~\bibnamefont {Nazarewicz}}, \bibinfo {author}
  {\bibfnamefont {E.}~\bibnamefont {Olsen}},\ and\ \bibinfo {author}
  {\bibfnamefont {O.~B.}\ \bibnamefont {Tarasov}},\ }\bibfield  {title}
  {\bibinfo {title} {Quantified limits of the nuclear landscape},\ }\href
  {https://doi.org/10.1103/PhysRevC.101.044307} {\bibfield  {journal} {\bibinfo
   {journal} {Physical Review C}\ }\textbf {\bibinfo {volume} {101}},\ \bibinfo
  {pages} {044307} (\bibinfo {year} {2020})}\BibitemShut {NoStop}%
\bibitem [{\citenamefont {Surman}\ \emph {et~al.}(2014)\citenamefont {Surman},
  \citenamefont {Mumpower}, \citenamefont {Cass}, \citenamefont {Bentley},
  \citenamefont {Aprahamian},\ and\ \citenamefont
  {McLaughlin}}]{surman_Sensitivity_2014}%
  \BibitemOpen
  \bibfield  {author} {\bibinfo {author} {\bibfnamefont {R.}~\bibnamefont
  {Surman}}, \bibinfo {author} {\bibfnamefont {M.}~\bibnamefont {Mumpower}},
  \bibinfo {author} {\bibfnamefont {J.}~\bibnamefont {Cass}}, \bibinfo {author}
  {\bibfnamefont {I.}~\bibnamefont {Bentley}}, \bibinfo {author} {\bibfnamefont
  {A.}~\bibnamefont {Aprahamian}},\ and\ \bibinfo {author} {\bibfnamefont
  {G.~C.}\ \bibnamefont {McLaughlin}},\ }\bibfield  {title} {\bibinfo {title}
  {Sensitivity studies for r-process nucleosynthesis in three astrophysical
  scenarios},\ }\href {https://doi.org/10.1051/epjconf/20146607024} {\bibfield
  {journal} {\bibinfo  {journal} {EPJ Web of Conferences}\ }\textbf {\bibinfo
  {volume} {66}},\ \bibinfo {pages} {07024} (\bibinfo {year}
  {2014})}\BibitemShut {NoStop}%
\bibitem [{\citenamefont {Mumpower}\ \emph {et~al.}(2015)\citenamefont
  {Mumpower}, \citenamefont {Surman}, \citenamefont {Fang}, \citenamefont
  {Beard},\ and\ \citenamefont {Aprahamian}}]{mumpower_Impact_2015}%
  \BibitemOpen
  \bibfield  {author} {\bibinfo {author} {\bibfnamefont {M.}~\bibnamefont
  {Mumpower}}, \bibinfo {author} {\bibfnamefont {R.}~\bibnamefont {Surman}},
  \bibinfo {author} {\bibfnamefont {D.~L.}\ \bibnamefont {Fang}}, \bibinfo
  {author} {\bibfnamefont {M.}~\bibnamefont {Beard}},\ and\ \bibinfo {author}
  {\bibfnamefont {A.}~\bibnamefont {Aprahamian}},\ }\bibfield  {title}
  {\bibinfo {title} {The impact of uncertain nuclear masses near closed shells
  on the r-process abundance pattern},\ }\href
  {https://doi.org/10.1088/0954-3899/42/3/034027} {\bibfield  {journal}
  {\bibinfo  {journal} {Journal of Physics G: Nuclear and Particle Physics}\
  }\textbf {\bibinfo {volume} {42}},\ \bibinfo {pages} {034027} (\bibinfo
  {year} {2015})}\BibitemShut {NoStop}%
\bibitem [{\citenamefont {Martin}\ \emph {et~al.}(2016)\citenamefont {Martin},
  \citenamefont {Arcones}, \citenamefont {Nazarewicz},\ and\ \citenamefont
  {Olsen}}]{martin_Impact_2016}%
  \BibitemOpen
  \bibfield  {author} {\bibinfo {author} {\bibfnamefont {D.}~\bibnamefont
  {Martin}}, \bibinfo {author} {\bibfnamefont {A.}~\bibnamefont {Arcones}},
  \bibinfo {author} {\bibfnamefont {W.}~\bibnamefont {Nazarewicz}},\ and\
  \bibinfo {author} {\bibfnamefont {E.}~\bibnamefont {Olsen}},\ }\bibfield
  {title} {\bibinfo {title} {Impact of {{Nuclear Mass Uncertainties}} on the
  r-process},\ }\href {https://doi.org/10.1103/PhysRevLett.116.121101}
  {\bibfield  {journal} {\bibinfo  {journal} {Physical Review Letters}\
  }\textbf {\bibinfo {volume} {116}},\ \bibinfo {pages} {121101} (\bibinfo
  {year} {2016})}\BibitemShut {NoStop}%
\bibitem [{\citenamefont {Mumpower}\ \emph {et~al.}(2016)\citenamefont
  {Mumpower}, \citenamefont {Surman}, \citenamefont {McLaughlin},\ and\
  \citenamefont {Aprahamian}}]{mumpower_Impact_2016}%
  \BibitemOpen
  \bibfield  {author} {\bibinfo {author} {\bibfnamefont {M.~R.}\ \bibnamefont
  {Mumpower}}, \bibinfo {author} {\bibfnamefont {R.}~\bibnamefont {Surman}},
  \bibinfo {author} {\bibfnamefont {G.~C.}\ \bibnamefont {McLaughlin}},\ and\
  \bibinfo {author} {\bibfnamefont {A.}~\bibnamefont {Aprahamian}},\ }\bibfield
   {title} {\bibinfo {title} {The impact of individual nuclear properties on
  r-process nucleosynthesis},\ }\href
  {https://doi.org/10.1016/j.ppnp.2015.09.001} {\bibfield  {journal} {\bibinfo
  {journal} {Progress in Particle and Nuclear Physics}\ }\textbf {\bibinfo
  {volume} {86}},\ \bibinfo {pages} {86} (\bibinfo {year} {2016})}\BibitemShut
  {NoStop}%
\bibitem [{\citenamefont {Mumpower}\ \emph {et~al.}(2017)\citenamefont
  {Mumpower}, \citenamefont {McLaughlin}, \citenamefont {Surman},\ and\
  \citenamefont {Steiner}}]{mumpower_Reverse_2017}%
  \BibitemOpen
  \bibfield  {author} {\bibinfo {author} {\bibfnamefont {M.~R.}\ \bibnamefont
  {Mumpower}}, \bibinfo {author} {\bibfnamefont {G.~C.}\ \bibnamefont
  {McLaughlin}}, \bibinfo {author} {\bibfnamefont {R.}~\bibnamefont {Surman}},\
  and\ \bibinfo {author} {\bibfnamefont {A.~W.}\ \bibnamefont {Steiner}},\
  }\bibfield  {title} {\bibinfo {title} {Reverse engineering nuclear properties
  from rare earth abundances in the r-process},\ }\href
  {https://doi.org/10.1088/1361-6471/44/3/034003} {\bibfield  {journal}
  {\bibinfo  {journal} {Journal of Physics G: Nuclear and Particle Physics}\
  }\textbf {\bibinfo {volume} {44}},\ \bibinfo {pages} {034003} (\bibinfo
  {year} {2017})}\BibitemShut {NoStop}%
\bibitem [{\citenamefont {Utama}\ and\ \citenamefont
  {Piekarewicz}(2017)}]{utamaRefiningMassFormulas2017}%
  \BibitemOpen
  \bibfield  {author} {\bibinfo {author} {\bibfnamefont {R.}~\bibnamefont
  {Utama}}\ and\ \bibinfo {author} {\bibfnamefont {J.}~\bibnamefont
  {Piekarewicz}},\ }\bibfield  {title} {\bibinfo {title} {Refining mass
  formulas for astrophysical applications: {{A Bayesian}} neural network
  approach},\ }\href {https://doi.org/10.1103/PhysRevC.96.044308} {\bibfield
  {journal} {\bibinfo  {journal} {Physical Review C}\ }\textbf {\bibinfo
  {volume} {96}},\ \bibinfo {pages} {044308} (\bibinfo {year}
  {2017})}\BibitemShut {NoStop}%
\bibitem [{\citenamefont {Foreman-Mackey}(2016)}]{corner}%
  \BibitemOpen
  \bibfield  {author} {\bibinfo {author} {\bibfnamefont {D.}~\bibnamefont
  {Foreman-Mackey}},\ }\bibfield  {title} {\bibinfo {title} {corner.py:
  Scatterplot matrices in python},\ }\href
  {https://doi.org/10.21105/joss.00024} {\bibfield  {journal} {\bibinfo
  {journal} {The Journal of Open Source Software}\ }\textbf {\bibinfo {volume}
  {1}},\ \bibinfo {pages} {24} (\bibinfo {year} {2016})}\BibitemShut {NoStop}%
\end{thebibliography}%
\end{document}


\title{Supplementary plots}


\begin{figure}
    \centering
    \includegraphics[scale=0.5,clip]{supplemental materials/figures/hessian_plot_E.pdf}
    \caption{Approximate Hessian matrix of energies with respect to Hamiltonian parameters. Large values indicate very sensitive parameters.}
    \label{fig:my_label}
\end{figure}

\begin{figure}
    \centering
    \includegraphics[scale=0.5,clip]{supplemental materials/figures/hessian_plot_GT.pdf}
    \caption{Approximate Hessian matrix of B(GT) with respect to Hamiltonian parameters. Large values indicate very sensitive parameters.}
    \label{fig:my_label}
\end{figure}

\begin{figure}
    \centering
    \includegraphics[scale=0.5,clip]{supplemental materials/figures/hessian_plot_E2.pdf}
    \caption{Approximate Hessian matrix of B(E2) with respect to Hamiltonian parameters. Large values indicate very sensitive parameters.}
    \label{fig:my_label}
\end{figure}

\begin{figure}
    \centering
    \includegraphics[scale=0.5,clip]{supplemental materials/figures/hessian_plot_M1.pdf}
    \caption{Approximate Hessian matrix of B(M1) with respect to Hamiltonian parameters. Large values indicate very sensitive parameters.}
    \label{fig:my_label}
\end{figure}

\begin{figure}
    \centering
    \includegraphics[scale=0.5,clip]{supplemental materials/figures/hessian_plot_SR(GT).pdf}
    \caption{Approximate Hessian matrix of the GT sum rule with respect to Hamiltonian parameters. Large values indicate very sensitive parameters.}
    \label{fig:my_label}
\end{figure}

\begin{figure}
    \centering
    \includegraphics[scale=0.5,clip]{supplemental materials/figures/hessian_plot_SR(E2).pdf}
    \caption{Approximate Hessian matrix of the E2 sum rule with respect to Hamiltonian parameters. Large values indicate very sensitive parameters.}
    \label{fig:my_label}
\end{figure}

\begin{figure}
    \centering
    \includegraphics[scale=0.5,clip]{supplemental materials/figures/hessian_plot_SR(M1).pdf}
    \caption{Approximate Hessian matrix of the M1 sum rule with respect to Hamiltonian parameters. Large values indicate very sensitive parameters.}
    \label{fig:my_label}
\end{figure}